\documentclass [12pt]{article}
\def\maj#1{\ifmmode\mbox{\usefont{U}{msb}{m}{n}#1}\else{\usefont{U}{msb}{m}{n}#1}\fi}
\def\v#1{\mathbf{#1}}
\usepackage[dvips]{graphics,epsfig,graphicx}
\usepackage{color}

\begin{document}

\title{\textbf{Excitons dressed by a sea of
excitons}}
\author{M. Combescot and O. Betbeder-Matibet
 \\ \small{\textit{GPS, Universit\'e Pierre et Marie
Curie and Universit\'e Denis Diderot,
CNRS,}}\\ \small{\textit{Campus Boucicaut, 140 rue de
Lourmel, 75015 Paris, France}}}
\date{}
\maketitle

\begin{abstract}
We here consider an exciton $i$ embedded in a
sea of $N$ identical excitons $0$. If the
excitons are bosonized, a bosonic
enhancement factor, proportional to $N$, is
found for $i=0$. If the 
exciton composite nature is kept, this enhancement not only exists
for $i=0$, but also for any exciton
having a center of mass momentum equal to the
sea exciton momentum. This physically
comes from the fact that an exciton with such a
momentum can be transformed into a
sea exciton by ``Pauli scattering'', \emph{i}.\
\emph{e}., carrier exchange with the
sea, making this $i$ exciton not so much
different from a $0$ exciton. This possible
scattering, directly linked to the composite nature of
the excitons, is irretrievably lost when the
excitons are bosonized.

This work in fact deals with the quite tricky scalar
products of $N$-exciton states. It actually constitutes a
crucial piece of our new many-body theory
for interacting composite bosons, because all physical effects
involving these composite bosons ultimately end by
calculating such scalar products. The ``Pauli
diagrams'' we here introduce to represent them, allow
to visualize many-body effects linked to carrier exchange in an easy way.
They are conceptually different from Feynman
diagrams, because of the special feature of the ``Pauli
scatterings'': These scatterings, which originate from
the departure from boson statistics, do not have their equivalent
in Feynman diagrams, the commutation rules for exact
bosons (or fermions) being included in the first line of the usual
many-body theories.
\end{abstract}

\vspace{2cm}


\newpage

\section{Introduction}

We are presently developing a
new many-body theory [1-7] able to handle
interactions between composite bosons --- like
the semiconductor excitons. The
development of such a theory is in fact
highly desirable, because, in the low density
limit, electron-hole pairs are known to form
bound excitons, so that, in this limit, to
manipulate excitons is surely a better idea than to manipulate free carriers.
However, the interaction
\emph{between} excitons is not an easy concept due to
carrier indistinguishability: Indeed, the
excitons, being made of two charged particles, of course interact through
Coulomb interactions. However, this Coulomb interaction can be
$(V_{ee'}+V_{hh'}-V_{eh'}-V_{e'h})$ or
$(V_{ee'}+V_{hh'}-V_{eh}-V_{e'h'})$ depending
if we see the excitons as 
$(e,h)$ and $(e',h')$, or $(e,h')$ and
$(e',h)$. In addition, excitons interact in a
far more subtle manner through Pauli
exclusion between their indistinguishable
components, \emph{in the absence of any Coulomb
process}. This ``Pauli interaction'' is
actually the novel and interesting part
of our new many-body theory for composite
bosons. It basically comes from the departure from boson
statistics, all previous theories, designed
for true bosons or true fermions, having the
corresponding commutation rules set up in the
first line [8]. In our theory, the fact that the
excitons are not exact bosons appears through
``Pauli scatterings'' $\lambda_{mnij}$
between the ``in'' excitons $(i,j)$ and the
``out'' excitons $(m,n)$. Their link to boson
departure is obvious from their definition.
Indeed, these Pauli scatterings appear
through [2,3]
\begin{equation}
[B_m,B_i^\dag]=\delta_{mi}-D_{mi}\ ,
\end{equation}
\begin{equation}
[D_{mi},B_j^\dag]=2\sum_n\lambda_{mnij}\,B_n
^\dag\ ,
\end{equation}
$B_i^\dag$ being the $i$ exciton creation
operator. It precisely reads in terms of the
exciton wave function $\phi_i(\v r_e,\v r_h)
=\langle\v r_e,\v r_h|B_i^\dag|v\rangle$ as
\begin{equation}
\lambda_{mnij}=\frac{1}{2}\int d\v r_e\,d\v
r_{e'}\,d\v r_h\,d\v r_{h'}\,\phi_m^\ast
(\v r_{e},\v r_h)\,\phi_n^\ast(\v r_{e'},\v
r_{h'})\,\phi_i(\v r_e,\v r_{h'})\,\phi_j(\v
r_{e'},\v r_{h})\ +\ (m\leftrightarrow n)\ .
\end{equation}
The above expression makes clear the fact
that $\lambda_{mnij}$ just corresponds to a
carrier exchange between two excitons (see
fig.\ 1a) without any Coulomb process, so
that $\lambda_{mnij}$ is actually a
dimensionless ``scattering''. It is possible
to show that for bound states,
$\lambda_{mnij}$ is of the order of
$\mathcal{V}_X/\mathcal{V}$, with
$\mathcal{V}_X$ being the exciton volume and
$\mathcal{V}$ the sample volume [6].

All physical quantities involving excitons
can be written as matrix elements of an
Hamiltonian dependent operator $f(H)$ between
$N$-exciton states, with usually most of them
in the ground state $0$. These
matrix elements formally read
\begin{equation}
\langle v|B_{m_1}\cdots
B_{m_n}B_0^{N-n}\,f(H)\,B_0^{\dag
N-n'}B_{i_1}^\dag\cdots
B_{i_{n'}}^\dag|v\rangle\ .
\end{equation}
They can be calculated by ``pushing'' $f(H)$
to the right in order to end with
$f(H)|v\rangle$, which is just
$f(0)|v\rangle$ if the vacuum is taken as the
energy origin. This push is done through a
set of commutations.
In the simplest case, $f(H)=H$, we have
\begin{equation}
HB_i^\dag=B_i^\dag(H+E_i)+V_i^\dag\ ,
\end{equation}
which just results from [2,3]
\begin{equation}
[H,B_i^\dag]=E_iB_i^\dag+V_i^\dag\ .
\end{equation}
We then push $V_i^\dag$ to the right
according to
\begin{equation}
[V_i^\dag,B_j^\dag]=\sum_{mn}\xi_{mnij}^\mathrm{dir}\,B_m^\dag
B_n^\dag\ ,
\end{equation}
to end with $V_i^\dag|v\rangle$ which is just $0$ due
to eq.\ (6) applied to $|v\rangle$.

Equations (6,7), along with eqs.\ (1,2), form the four key
equations of our many-body theory for interacting composite
excitons. $\xi_{mnij}^\mathrm{dir}$ is the second scattering of
this theory. It transforms the $(i,j)$ excitons into $(m,n)$
states, due to Coulomb processes \emph{between} them, as obvious
from its explicit expression:
\begin{eqnarray}
\xi_{mnij}^\mathrm{dir}=\frac{1}{2}\int d\v r_e\,d\v r_{e'}\,d\v
r_h\,d\v r_{h'}\,\phi_m^\ast(\v r_e,\v r_h)\,\phi_n^\ast(\v
r_{e'},\v r_{h'})\hspace{4cm}\nonumber\\
\times (V_{ee'}+V_{hh'}-V_{eh'}-V_{e'h})\,\phi_i(\v r_e,\v
r_h)\,\phi_j(\v r_{e'},\v r_{h'})\ +(m\leftrightarrow n)\ .
\end{eqnarray}
Note that, in $\xi_{mnij}^\mathrm{dir}$, the ``in'' and ``out''
excitons are made with the same pairs, while, in
$\lambda_{mnij}$, they have exchanged their carriers. Due to
dimensional arguments, these $\xi_{mnij}^\mathrm{dir}$ for bound
states are of the order of $R_X\mathcal{V}_X/\mathcal{V}$, with
$R_X$ being the exciton Rydberg [6].

Another $f(H)$ of interest can be $1/(a-H)$, with $a$ possibly
equal to $(\omega+i\eta)$ as in problems involving
photons. In order to push $1/(a-H)$ to the right, we can
use [4]
\begin{equation}
\frac{1}{a-H}\,B_i^\dag=B_i^\dag\,
\frac{1}{a-H-E_i}+\frac{1}{a-H}\,V_i^\dag\,\frac{1}{a-H-E_i}\ ,
\end{equation}
which follows from eq.\ (5). In pushing $1/(a-H)$ to the
right, we generate Coulomb terms through the $V_i^\dag$ part of
eq.\ (9). Due to dimensional arguments, these terms ultimately
read as an expansion in 
$\xi_{mnij}^\mathrm{dir}$ over an energy denominator which can
be either a detuning or just a difference between exciton
energies, depending on the problem at hand.

A last $f(H)$ of interest is $e^{-iHt}$ which appears in
problems involving time evolution. In order to push $e^{-iHt}$
to the right, we can use [7]
\begin{equation}
e^{-iHt}\,B_i^\dag=B_i^\dag\,e^{-i(H+E_i)t}+W_i^\dag(t)\ ,
\end{equation}
\begin{equation}
W_i^\dag(t)=-\int_{-\infty}^{+\infty}\frac{dx}{2i\pi}\,
\frac{e^{-i(x+i\eta)t}}{x-H+i\eta}\,V_i^\dag\,\frac{1}{x-H-E_i
+i\eta}\ .
\end{equation}
Equations (10,11) result from the integral representation of the
exponential, namely
\begin{equation}
e^{-iHt}=-\int_{-\infty}^{+\infty}\frac{dx}{2i\pi}\,
\frac{e^{-i(x+i\eta)t}}{x-H+i\eta}\ ,
\end{equation}
valid for $t$ and $\eta$ positive, combined with eq.\ (9). Again
additional Coulomb terms appear in passing $e^{-iHt}$ over
$B_i^\dag$.

By comparing eqs.\ (5,9,10), we see that, when we pass $f(H)$
over $B_i^\dag$, we essentially replace it by $f(H+E_i)$,
as if the $i$ exciton was not interacting with the other
excitons, within a Coulomb term which
takes care of these interactions, $f(H+E_i)$ being in some sense
the zero order contribution of $f(H)$.

Once we have pushed all the $H$'s up to $|v\rangle$ and
generated very many Coulomb scatterings
$\xi_{mnij}^\mathrm{dir}$, we end with scalar products of
$N$-exciton states which look like eq.\ (4) with $f(H)=1$. Then,
we start to push the $B$'s to the right according to eqs.\
(1,2), to end with $B|v\rangle$ which is just zero. This set of
pushes now makes appearing the Pauli scatterings
$\lambda_{mnij}$. 

In the case of $N=2$, eqs.\ (1,2) readily give
the scalar product of two-exciton states as [2]
\begin{equation}
\langle v|B_m\,B_n\,B_i^\dag\,B_j^\dag|v\rangle=\delta_{mi}\,
\delta_{nj}+\delta_{mj}\,\delta_{ni}-2\lambda_{mnij}\ .
\end{equation}
For large $N$, the calculation of similar scalar products
is actually very tricky. We expect them to depend on $N$
and to contain many $\lambda_{mnij}$'s. 

The $N$ dependence of these scalar products is
in fact crucial because physical quantities
must ultimately depend on $\eta=N\mathcal{V}_X/\mathcal{V}$,
with $\mathcal{V}_X/\mathcal{V}$ coming either from Pauli
scatterings or from Coulomb scatterings --- with possibly an
additional factor
$N$, if we look for something extensive.
However, as the scalar products of
$N$-exciton states are not physical quantities, they can very
well contain superextensive terms in $N^p\eta^n$ which
ultimately disappear from the final expressions of the physical
quantities. To handle these factors $N$ properly --- and
their possible cancellations --- is thus crucial.

In previous works [1,5], we have calculated the simplest of
these scalar products of $N$-exciton states, namely $\langle v|
B_0^NB_0^{\dag N}|v\rangle$. We found it equal to $N!$, as
for exact bosons, multiplied by a corrective factor $F_N$
which comes from the fact that excitons are composite bosons,
\begin{equation}
\langle v|B_0^N\,B_0^{\dag N}|v\rangle=N!\,F_N\ .
\end{equation}
This $F_N$ factor is actually superextensive, since it behaves
as $e^{-N\,O(\eta)}$ (see ref.\ [5]). In large enough
samples, $N\eta$ can be extremely large even for $\eta$
small, which makes $F_N$
exponentially small. In physical quantities however,
$F_N$ never appears alone, but through ratios like
$F_{N-p}/F_N$ which actually read as
$1+O(\eta)$ for $p\ll N$. This restores the expected $\eta$
dependence of these physical quantities.

The present paper in fact deals with determining the interplay
between the possible factors $N$ and the various $\lambda$'s
which appear in scalar products of $N$-exciton states. These
Pauli scatterings $\lambda$ being the original part of our many-body theory
for interacting composite bosons, the understanding of this
interplay is actually fundamental to master many-body effects
between excitons at any order in $\eta=N\mathcal{V}_X/\mathcal
{V}$. This will in particular allow us to cleanly show the
cancellation of superextensive terms which can possibly appear
in the intermediate stages of the theory [7]. 

This paper can appear as somewhat formal. It however constitutes
one very important piece of this new many-body theory, because
all physical effects between $N$ interacting excitons ultimately
end with calculating such scalar products. Problems involving two
excitons only[7] are in fact rather simple to solve because they
only need the scalar product of two-exciton states given in eq.\
(13). The real challenging difficulty which actually remains in
order to handle many-body effects between $N$ excitons at any order in $\eta$,
is to produce the equivalent of eq.\ (13) for large $N$.

In usual many-body effects, Feynman diagrams [8] have been proved to
be quite convenient to understand the physics of interacting
fermions or bosons. We can expect the introduction of diagrams
to be also quite convenient to understand the physics of
interacting composite bosons. It is however clear that diagrams
representing carrier exchange between 
$N$ excitons have to be conceptually new: In
them, must enter the Pauli scatterings which take care of the
departure from boson statistics. As the fermion or boson
statistics is included in the first lines of the usual many-body
theories, the Pauli scatterings do not have their equivalent in
Feynman diagrams. Another important part of the present paper is
thus to present these new ``Pauli diagrams'' and to derive some of
their specific rules. As we will show, these Pauli diagrams are
in fact rather tricky because they can look rather differently
although they represent exactly the same quantity. To
understand why these different diagrams are indeed equivalent,
is actually crucial to master these Pauli diagrams. This is the
subject of the last section of this paper. It goes through the
introduction of ``exchange skeletons'' which are the basic
quantities for carrier exchanges between more than two excitons.
Their appearance is physically reasonable because Pauli
exclusion is $N$-body ``at once'', so that, when it plays a
role, it in fact correlates all the carriers of the involved
excitons through a unique process, even if this process can be
decomposed into exchanges between two excitons only, as in the
Pauli scatterings $\lambda_{mnij}$.

From a technical point of view, it is of course possible to
calculate the scalar products of
$N$-exciton states, just through blind algebra based on eqs.\
(1,2), and to get the right answer. However, in order to
understand the appearance of the extra factors $N$ which go in
front of the ones in
$N\lambda\simeq N\mathcal{V}_X/\mathcal{V}=\eta$ and which are
crucial to ultimately withdraw superextensive terms from physical
quantities, it is in fact convenient to introduce the
concept of \emph{``excitons dressed by a sea of excitons''}, because
these extra factors $N$ are physically linked to the underlying
bosonic character of the excitons which is enhanced by the
presence of an exciton sea. We will show that these extra
factors $N$ are linked to the topology of the
diagrammatic representation of these scalar products, which
appears as ``disconnected'' when these extra $N$'s exist. This is
after all not very astonishing because disconnected Feynman
diagrams are known to also generate superextensive terms.

Let us introduce the exciton $i$ dressed by a sea of $N$ excitons
$0$ as
\begin{equation}
|\psi_i^{(N)}\rangle=\frac{B_0^NB_0^{\dag N}}{\langle v|B_0^N
B_0^{\dag N}|v\rangle}\ B_i^\dag|v\rangle\ ,
\end{equation}
The denominator $\langle v|B_0^N
B_0^{\dag N}|v\rangle$ is a normalization factor which
makes the operator in front of $B_i^\dag$ appearing as an
identity in the absence of Pauli interactions with the exciton
sea. Indeed, we can check that the vacuum state, dressed in the
same way as
\begin{equation}
|\psi^{(N)}\rangle=\frac{B_0^NB_0^{\dag N}}{\langle v|B_0^N
B_0^{\dag N}|v\rangle}\ |v\rangle\ ,
\end{equation}
is just $|v\rangle$, as expected because no interaction can exist
with the vacuum. On the opposite,
subtle Pauli interactions take place between the exciton sea
and an additional exciton $i$. As
the dressed exciton $i$
contains one more $B^\dag$ than the number of $B$'s, it
is essentially a one-pair state. It can be written either in
terms of free electrons and holes, or better in terms of
one-exciton states. Since these states are the one-pair
eigenstates of the semiconductor Hamiltonian $H$, they obey the
closure relation $1=\sum_m B_m^\dag|v\rangle\langle v|B_m$. So
that
$|\psi_i^{(N)}\rangle$ can be written as
\begin{equation}
|\psi_i^{(N)}\rangle=\sum_mA_N(m,i)\,B_m^\dag|v\rangle\ ,
\end{equation}
\begin{equation}
A_N(m,i)=\frac{\langle v|B_m\,B_0^N\,B_0^{\dag
N}\,B_i^\dag|v\rangle}{\langle v|B_0^N\,B_0^{\dag N}|v\rangle}
\ .
\end{equation}
This decomposition of $|\psi_i^{(N)}\rangle$ on one-exiton
states makes appearing the scalar product of $(N+1)$-exciton
states with $N$ of them in the exciton sea. 

As the
physics which controls the extra factors $N$ in these scalar
products, is actually linked to the underlying bosonic character
of the excitons, let us first consider boson-excitons in order to
see how a sea of $N$ boson-excitons $0$ affects them.

\section{Boson-excitons dressed by a sea of excitons}

Instead of eq.\ (1), the commutation rule for boson-excitons is
$[\bar{B}_m,\bar{B}_i^\dag]=\delta_{mi}$, so that the
deviation-from-boson operator $D_{mi}$ for boson-excitons is
zero, as the Pauli scatterings $\lambda_{mnij}$. From this boson
commutator, we get by induction
\begin{equation}
[\bar{B}_0^N,\bar{B}_i^\dag]=\bar{B}_0^{N-1}[\bar{B}_0,\bar{B}_i
^\dag]+[\bar{B}_0^{N-1},\bar{B}_i^\dag]\bar{B}_0=N\delta_{0i}\,
\bar{B}_0^{N-1}\ .
\end{equation}
So that $\bar{B}_0^N\bar{B}_0^{\dag N}|v\rangle=N!\,|v\rangle$,
which shows that the normalization factor $F_N$ is just 1 for
boson-excitons, while $|\bar{\psi}_0^{(N)}\rangle=(N+1)\bar{B}_0
^\dag|v\rangle$ and
$|\bar{\psi}_{i\neq 0}^{(N)}\rangle=\bar{B}_i
^\dag|v\rangle$. So that, we do have
\begin{equation}
|\bar{\psi}_i^{(N)}\rangle=(N\delta_{i0}+1)\,\bar{B}_i^\dag
|v\rangle\ .
\end{equation}
The factor $N$ which appears in this equation is physically
linked to the well known bosonic enhancemen [9]. The memory
of such an effect must \emph{a priori} exist for composite
bosons, such as the excitons. However, subtle changes are
expected due to their underlying fermionic character. Let us now
see how this bosonic enhancement, obvious for
boson-excitons, does appear for exact excitons.

\section{Exact excitons dressed by a sea of excitons}

The commutation rule for exact excitons is given in eq.\ (1). By
taking another commutation, we generate eq.\ (2) which defines
the Pauli scattering $\lambda_{mnij}$. From it, we easily get by
induction
\begin{equation}
[D_{mi},B_0^{\dag N}]=2N\sum_n\lambda_{mn0i}\,B_n^\dag B_0^{\dag
N-1}\ ,
\end{equation}
its conjugate leading to
\begin{equation}
[B_0^N,D_{mi}]=2N\sum_j\lambda_{m0ji}\,B_0^{N-1}B_j\ ,
\end{equation}
since $D_{mi}^\dag=D_{im}$, while $\lambda_{mnij}^\ast=\lambda_
{ijmn}$. Equation (21) allows to generalize eq.\ (1) as
\begin{equation}
[B_m,B_0^{\dag N}]=NB_0^{\dag N-1}(\delta_{m0}-D_{m0})-N(N-1)
\sum_n\lambda_{mn00}B_n^\dag B_0^{\dag N-2}\ ,
\end{equation}
its conjugate leading to
\begin{equation}
[B_0^N,B_i^\dag]=N(\delta_{0i}-D_{0i})B_0^{N-1}-N(N-1)\sum_j
\lambda_{00ij}B_jB_0^{N-2}\ .
\end{equation}
(We can note that eq.\ (19) for bosons just follows from eq.\
(24), since the deviation-from-boson operator
$D_{mi}$ and the Pauli scattering $\lambda_{mnij}$ are equal to
zero in the case of boson-excitons).

In order to grasp the bosonic enhancement for exact
excitons, let us start with the ``best case'' for such an
enhancement, namely an exciton $0$ dressed by a sea of $N$
excitons $0$.

\subsection{Exciton $0$ dressed by $N$ excitons $0$}

According to eq.\ (17), this dressed exciton can be writen as
\begin{equation}
|\psi_0^{(N)}\rangle=(N+1)\sum_m\zeta_N(m)B_m^\dag|v\rangle\,
\end{equation}
in which we have set
\begin{equation}
\zeta_N(m)=\frac{A_N(m,0)}{N+1}=\frac{\langle v|B_0^NB_mB_0
^{\dag N+1}|v\rangle}{(N+1)!F_N}\ .
\end{equation}
This $\zeta_N(m)$, which is just $F_{N+1}/F_N\simeq 1+O(\eta)$
for
$m=0$, will appear to be a quite useful quantity in the
following. To calculate it, we rewrite $B_mB_0^{\dag N+1}$
according to eq.\ (23). Since $D_{m0}|v\rangle=0$, which follows
from eq.\ (1) applied to $|v\rangle$, this readily gives the
recursion relation between the $\zeta_N(m)$'s as
\begin{equation}
\zeta_N(m)=\delta_{m0}-\frac{F_{N-1}}{F_N}\,N\sum_n
\lambda_{mn00}\,\zeta_{N-1}^\ast(n)\ .
\end{equation}
Its diagrammatic representation is shown in fig.\ 2a as well
as its iteration (fig.\ 2b). Its solution is
\begin{equation}
\zeta_N(m)=\sum_{p=0}^N (-1)^p\frac{F_{N-p}}{F_N}\,\frac{N!}
{(N-p)!}\,z^{(p)}(m,0)\ ,
\end{equation}
with $z^{(0)}(m,0)=\delta_{m0}$, while
\begin{equation}
z^{(p)}(m,0)=\sum_n\lambda_{mn00}\,\left[z^{(p-1)}(n,0)\right]
^\ast\ .
\end{equation}
$z^{(p)}(m,0)$ can be represented by a diagram with $(p+1)$
lines, the lowest one being $(m,0)$, while the $p$ other lines
are
$(0,0)$. These lines are connected by $p$ Pauli scatterings
which are in zigzag, alternatively right, left, right\ldots (see
fig.\ 2b). For
$p=1$, $z^{(p)}(m,0)$ is just $\lambda_{m000}$, while for $p=2$,
it reads
$\sum_n\lambda_{mn00}\lambda_{00n0}$ and so on \ldots Fig.\ 
2c also shows $\zeta_N^\ast(i)$, easy to obtain from
$\zeta_N(m)$ by noting that
$\lambda_{mnij}^\ast=\lambda_{ijmn}$, so that $\zeta_N^\ast(i)$
is just obtained from $\zeta_N(i)$ by a symmetry right-left:
In $\zeta_N^\ast(i)$, the zigzag in fact appears  as left, right,
left,\ldots  

Since we are ultimately interested in possible extra factors
$N$, it can be of interest to understand the appearance of
$N$'s in
$\zeta_N(m)$. If we forget about the composite nature of the
exciton, \emph{i}.\ \emph{e}., if we drop all carrier
exchanges with the exciton sea, the electron and hole of the
exciton $0$ are tight for ever as for boson-excitons, so that we
should have for $|\psi_0^{(N)}\rangle$ the same result as the
one for boson-excitons, namely $|\psi_0^{(N)}\rangle\simeq
(N+1)B_0^\dag|v\rangle$. This leads to $\zeta_N(m)\simeq
\delta_{m0}$ at lowest order in $\lambda$. The composite exciton
$0$ can however exchange its electron or its hole with one sea
exciton to become an $m$ exciton. Since there are $N$ possible
excitons in the sea for such an exchange, the first order term
in exchange scattering must appear with a factor $N$. Another
exciton, among the $(N-1)$ left in the sea, can also participate
to these carrier exchanges so that the second order term in
Pauli scattering must appear with a $N(N-1)$ prefactor; and
so on \ldots

From this iteration, we thus conclude that $\zeta_N(m)$ contains
the same number of factors $N$ as the number of $\lambda$'s.
Since, for $0$ and $m$ being bound states, these $\lambda$'s are
in $\mathcal{V}_X/\mathcal{V}$, while $F_{N-p}/F_N$ reads as an
expansion in $\eta$ (see ref.\ [5]), we thus find that, in the
large $N$ limit, $\zeta_N(m)$ can be written as an expansion in
$\eta$, without any extra factor $N$ in front. This thus shows
that $|\psi_0^{(N)}\rangle$ contains the same bosonic enhancement
factor $(N+1)$ as the one of dressed boson-excitons
$|\bar{\psi}_0^{(N)}\rangle$. Let us however stress that the
relative weight of the $B_0^\dag|v\rangle$ state in 
$|\psi_0^{(N)}\rangle$, namely $\zeta_N(0)$, which is exactly 1
in the case of boson-excitons, is somewhat smaller than 1 due to
possible carrier exchanges with the exciton sea. From the
iteration of $\zeta_N(0)$, we find that this weight reads
$\zeta_N(0)=1-(F_{N-1}/F_N)N\lambda_{0000}+\cdots$, which is
nothing but $F_{N+1}/F_N$ as can be directly seen from eq.\ (26).

To compensate this decrease of the $B_0^\dag|v\rangle$ state
weight, $|\psi_0^{(N)}\rangle$ has non-zero components on the
other exciton states $B_{m\neq 0}^\dag|v\rangle$, in contrast
with the boson-exciton case. We can however note that, since
$\zeta_N(m)=0$ for $\v Q_m\neq \v Q_0$, due to momentum
conservation in Pauli scatterings, the other exciton
states making $|\psi_0^{(N)}\rangle$ must have the same
momentum $\v Q_0$ as the one of the $0$ excitons.

We thus conclude that one exciton $0$ dressed by a sea of $N$
excitons $0$, exhibits the same enhancement factor $(N+1)$ as
the one which appears for boson-excitons. This dressed exciton
however has additional components on other exciton states which
have a momentum equal to the $0$ exciton momentum $\v Q_0$.

The existence of such a bosonic enhancement for the
exciton
$0$ can appear as somewhat normal because excitons are, after
all, not so far from real bosons. We will now show that a similar
enhancement, \emph{i}.\ \emph{e}., an additional
prefactor $N$, also exists for excitons different from $0$ but
having a center of mass momentum equal to $\v Q_0$. Before
showing it from hard algebra, let us physically explain why this
has to be expected: From the two possible ways to form two
excitons out of two electron-hole pairs, we have shown that
\begin{equation}
B_i^\dag B_j^\dag=-\sum_{mn}\lambda_{mnij}\,B_m^\dag B_n^\dag\ .
\end{equation}
$B_{i\neq 0}^\dag B_0^\dag$ can thus be written as a sum of an
exciton $m$ and an exciton $n$ with $\v Q_m+\v Q_n=\v Q_i+\v
Q_0$ due to momentum conservation included in the Pauli
scatterings $\lambda_{mnij}$. This shows that, for an exciton
$i$ with $\v Q_i=\v Q_0$, $B_{i\neq 0}^\dag B_0^\dag$ has a
non-zero contribution on $B_0^{\dag 2}$, so that this $i\neq 0$
exciton, in the presence of other excitons $0$, is partly
a $0$ exciton. A bosonic enhancement has thus to exist
for any exciton $i$ with $\v Q_i=\v Q_0$.

\subsection{Exciton $i$ dressed by $N$ excitons $0$}

Let us now consider an exciton with arbitrary $i$. We will show
that there are essentially two kinds of such excitons, the ones
with $\v Q_i=\v Q_0$ and the ones with $\v Q_i\neq \v Q_0$: Since
a
$\v Q_i=\v Q_0$ exciton can be transformed into an $i=0$ exciton
by carrier exchange with the exciton sea, it is clear that the
excitons with $\v Q_i\neq \v Q_0$ are in fact the only
ones definitely different from $0$ excitons; this is why they
should be dressed differently.

The exciton $i$ dressed by $N$ excitons $0$ reads as eq.\ (17).
From eq.\ (26), we already know that $A_N(0,i)$ is just
$(N+1)\zeta_N(i)$, so that we are left with determining the
scalar product $A_N(m,i)$ for $(m,i)\neq 0$.

There are many ways to calculate $A_N(m,i)$: We can for example
start with $[B_m,B_i^\dag]$ given in eq.\ (1), or with
$[B_m,B_0^{\dag N}]$ given in eq.\ (23), or even with
$[B_0^N,B_i^\dag]$ given in eq.\ (24). While these last two
commutators lead to calculations essentially equivalent, the
first one may appear somewhat better at first, because it does
not destroy the intrinsic $(m,i)$ symmetry of the $A_N(m,i)$
matrix element. These various ways to
calculate $A_N(m,i)$ must of course end by giving exactly the
same result. However, it turns out that the diagrammatic
representations of $A_N(m,i)$ these various ways generate, look
at first rather different. We will, in this section, present the
calculation of $A_N(m,i)$ which leads to the ``nicest''
diagrams, \emph{i}.\ \emph{e}., the ones which are the easiest
to memorize. We leave the discussion of the other diagrammatic
representations of $A_N(m,i)$ and their equivalences for the
last part of this work.

\subsubsection{Recursion relation between $A_N(m,i)$ and
$A_{N-2}(n,i)$}

We start with $[B_m,B_i^\dag]$ given in eq.\ (1). This leads to
write $A_N(m,i)$ as
\begin{equation}
A_N(m,i)=a_N(m,i)+\hat{A}_N(i,m)\ ,
\end{equation}
in which we have set
\begin{equation}
a_N(m,i)=\delta_{mi}-\langle v|B_0^N\,D_{mi}\,B_0^{\dag
N}|v\rangle/N!F_N\ ,
\end{equation}
\begin{equation}
\hat{A}_N(i,m)=\langle v|B_0^NB_i^\dag B_mB_0^{\dag N}|v
\rangle/N!F_N\ .
\end{equation}

To calculate the matrix element appearing in $a_N(m,i)$, we can
either use $[D_{mi},B_0^{\dag N}]$ given in eq.\ (21), or
$[B_o^N,D_{mi}]$ given in eq.\ (22). With the first choice, we
find
\begin{equation}
a_N(m,i)=\delta_{mi}-2\frac{F_{N-1}}{F_N}\,N\sum_j\lambda_{mj0i}
\,\zeta_{N-1}^\ast(j)\ .
\end{equation}
Fig.\ 3a shows the diagrammatic representation of eq.\
(34), while fig.\ 3b shows the corresponding expansion of
$a_N(m,i)$ deduced from the
diagrammatic expansion of $\zeta_N^\ast(i)$ given in fig.\ 2c.
By injecting eq.\ (28) giving $\zeta_N(m)$ into eq.\ (34), we
find 
\begin{equation}
a_N(m,i)=\delta_{mi}+2\sum_{p=1}^N(-1)^p\,\frac{F_{N-p}}{F_N}\,\frac{N!}
{(N-p)!}\,z^{(p)}(m,i)\ ,
\end{equation}
where $z^{(p)}(m,i)$ is such that
\begin{equation}
z^{(p)}(m,i)=\sum_j\lambda_{mj0i}\,z^{(p-1)\ast}(j,0)\ .
\end{equation}
As shown in fig.\ 3b, $z^{(p)}(m,i)$ is a
zigzag diagram like
$z^{(p)}(m,0)$, with the lowest line $(m,0)$ replaced by
$(m,i)$.

If we now turn to $\hat{A}_N(i,m)$, there are \emph{a priori} two
ways to calculate it: Either we use $[B_m,B_0^{\dag N}]$, or we
use $[B_0^N,B_i^\dag]$. However, if we want to write 
$\hat{A}_N(i,m)$ in terms of $A_{N-2}(n,i)$, we must keep
$B_i^\dag$ so that $[B_0^N,B_i^\dag]$ is not appropriate.
Equation (23) then leads to
\begin{equation}
\hat{A}_N(i,m)=\frac{F_{N-1}}{F_N}\,N\,\delta_{m0}\,
\zeta_{N-1}^\ast(i)-\frac{N(N-1)}{N!F_N}\sum_j\lambda_{mj00}\,
\langle v|B_0^NB_i^\dag B_j^\dag B_0^{\dag N-2}|v\rangle\ .
\end{equation}
The above matrix element can be calculated either with
$[B_0^N,B_j^\dag]$ or with $[B_0^N,B_i^\dag]$. From the first
commutator --- which is the one which allows to keep $B_i^\dag$
--- we get
\begin{equation}
\hat{A}_N(i,m)=N\,b_N(m,i)+\frac{F_
{N-2}}{F_N}\,N(N-1)\sum_{nj}\lambda_{mj00}\lambda_{00nj}\,
A_{N-2}(n,i)\ ,
\end{equation}
in which we have set
\begin{equation}
b_N(m,i)=\frac{F_{N-1}}{F_N}\,\delta_{m0}\,\zeta_{N-1}^\ast(i)
-\frac{F_{N-2}}{F_N}
\,(N-1)\,\lambda_{m000}\,\zeta_{N-2}^\ast(i)\ .
\end{equation}
By using the expansion of $\zeta_N(m)$ given in eq.\ (28), this
equation leads to write $b_N(m,i)$ as
\begin{eqnarray}
b_N(m,i)=\frac{F_{N-1}}{F_N}\,\delta_{m0}\,\delta_{0i}+\sum_{p=1}
^{N-1}(-1)^p\,\frac{(N-1)!}{(N-1-p)!}\,\frac{F_{N-1-p}}{F_N}
\hspace{3cm}\nonumber \\ \times
\left[z^{(0)}(m,0)\,z^{(p)\ast}
(i,0)+z^{(1)}(m,0)\,z^{(p-1)\ast}(i,0)\right]\ .
\end{eqnarray}

The diagrammatic representation of eq.\ (39) is shown in fig.\
4a. From it and the diagrams of fig.\ 2c for
$\zeta_N^\ast(i)$, we obtain the Pauli expansion of $b_N(m,i)$
shown in fig.\ 4b. It is just the diagrammatic representation
of eq.\ (40). We see that
$b_N(m,i)$ is made of diagrams which can be cut into two pieces.
We also see that, while
$a_N(m,i)$ differs from 0 for
$\v Q_m=\v Q_i$ only, due to momentum conservation included in
the Pauli scatterings, we must have $\v Q_m=\v Q_i=\v Q_0$ to
have $b_N(m,i)\neq 0$, since $\zeta_N(m)$ is 0 for $\v Q_m\neq
\v Q_0$, as previously shown.

From eqs.\ (31) and (38), we thus find that 
$A_N(m,i)$ obeys the recursion relation
\begin{equation}
A_N(m,i)=a_N(m,i)+N\,b_N(m,i)+\frac{F_{N-2}}
{F_N}\,N(N-1)\sum_{nj}\lambda_{mj00}\lambda_{00nj}\,A_{N-2}(n,i)
\ ,
\end{equation}
with $b_N(m,i)=0$ if $\v Q_m=\v Q_i\neq \v Q_0$.

\subsubsection{Determination of $A_N(m,i)$ using $A_{N-2}(n,i)$}

From the fact that we just need to have $\v Q_m=\v
Q_i$ for $a_N(m,i)\neq 0$, while $b_N(m,i)\neq 0$
imposes
$\v Q_m=\v Q_i=\v Q_0$, we are led to divide $A_N(m,i)$ into a
contribution which exists whatever the $i$ exciton momentum is
and a contribution which only exists when $\v Q_i$ is equal
to the sea exciton momentum
$\v Q_0$. This gives
\begin{equation}
A_N(m,i)=\alpha_N(m,i)+N\,\beta_N(m,i)\ ,
\end{equation}
where $\alpha_N(m,i)$ and $\beta_N(m,i)$ obey the two
recursion relations
\begin{equation}
\alpha_N(m,i)=a_N(m,i)+\frac{F_{N-2}}{F_N}\,N(N-1)\sum_{nj}
\lambda_{mj00}\lambda_{00nj}\,\alpha_{N-2}(n,i)\ ,
\end{equation}
\begin{equation}
\beta_N(m,i)=b_N(m,i)+\frac{F_{N-2}}{F_N}\,(N-1)(N-2)
\sum_{nj}
\lambda_{mj00}\lambda_{00nj}\,\beta_{N-2}(n,i)\ .
\end{equation}

\vspace{1cm}

\noindent\textbf{a) Part of $A_N(m,i)$ which exists whatever $\v
Q_i(=\v Q_m)$ is}

The part of $A_N(m,i)$ which exists for any exciton $i$ is
$\alpha_N(m,i)$. The diagrammatic representation of its
recursion relation (43) is shown in fig.\ 5a, as well
as its iteration (fig.\ 5b). If, in it, we insert the
diagrammatic representation of $a_N(m,i)$ given in fig.\ 3b,
we end with the diagrammatic representation of $\alpha_N(m,i)$
shown in fig.\ 5c. Note that we have used
$\lambda_{mn0i}=\lambda_{mni0}$ in order to get rid of the
factor 2 appearing in $a_N(m,i)$. Using eq.\ (35) for $a_N(m,i)$,
it is easy to check that the solution of eq.\ (43) reads
\begin{equation}
\alpha_N(m,i)=\sum_{p=0}^N(-1)^p\,\frac{F_{N-p}}{F_N}\,\frac{N!}
{(N-p)!}\,Z^{(p)}(m,i)\ ,
\end{equation}
where $Z^{(p)}(m,i)$  obeys the recursion relation
\begin{equation}
Z^{(p)}(m,i)=\hat{z}^{(p)}(m,i)+\sum_{nj}\lambda_{mj00}\,
\lambda_{00nj}\,Z^{(p-2)}(n,i)\ ,
\end{equation}
with $\hat{z}^{(0)}=\delta_{mi}$, while $\hat{z}^{(p\neq 0)}(m,i)
=2z^{(p)}(m,i)$. In agreement with fig.\ 5c, this leads to
represent
$Z^{(p)}(m,i)$ as a sum of zigzag diagrams with
$p$ Pauli scatterings, located alternatively right, left,
right,\ldots, the index $m$ being always at the left bottom,
while $i$ can be at all possible places on the right.
$Z^{(p)}(m,i)$ thus contains $(p+1)$ diagrams which reduce to
one, namely
$Z^{(0)}(m,i)=\delta_{mi}$, when $p=0$.

From fig.\ 5c, we also see that $\alpha_N(m,i)$ contains as
many $N$'s as $\lambda$'s so that it ultimately depends on
$(N,\lambda)$'s through $\eta$.

\vspace{1cm}

\noindent\textbf{b) Part of $A_N(m,i)$ which exists for $\v
Q_i(=\v Q_m)=\v Q_0$ only}

The part of $A_N(m,i)$ which only exists when the $i$ and
$m$ excitons have the same momentum as the sea exciton one, is 
$N\,\beta_N(m,i)$. The diagrammatic
representation of the recursion relation (44) for $\beta_N$
is shown in fig.\ 6a, as well as its iteration (fig.\
6b). Using eq.\ (40) for $b_N(m,i)$ and  eq.\ (29), it is easy
to check that the solution of the recursion relation (44) reads
\begin{equation}
\beta_N(m,i)=\sum_{p=0}^{N-1}(-1)^p\,\frac{F_{N-1-p}}{F_N}\,
\frac{(N-1)!}{(N-1-p)!}\,\sum_{q=0}^pz^{(q)}(m,0)\,z^{(p-q)\ast}
(i,0)\ .
\end{equation}
This is exactly what we get if, in the expansion of
$\beta_N(m,i)$ in terms of
$b_N(n,i)$ shown in fig.\ 6b, we insert the expansion of
$b_N(n,i)$ in terms of Pauli scatterings
shown in fig.\ 4b, (see fig.\ 6c):
The diagrams making $\beta_N(m,i)$ are thus made of two
pieces, in agreement with eq.\ (47). We also see that
$\beta_N(m,i)$ contains as many
$N$'s as $\lambda$'s so that $\beta_N(m,i)$, like
$\alpha_N(m,i)$, is an $\eta$ function. 

\vspace{1cm}

\noindent\textbf{c) $N$ dependence of $A_N(m,i)$}

If we now come back to the expression (41) for $A_N(m,i)$, we
see that when $\beta_N(m,i)=0$, \emph{i}.\ \emph{e}., when $\v
Q_m=\v Q_i\neq \v Q_0$, the $N$'s in $A_N(m,i)$ simply appear
through products $N\lambda$. On the opposite, $A_N(m,i)$
contains an extra prefactor $N$ when $\beta_N(m,i)\neq 0$,
\emph{i}.\ \emph{e}., when $\v Q_m=\v Q_i=\v Q_0$: This extra $N$
is the memory of the bosonic enhancement found for the dressed
exciton $i=0$. As already explained above, this bosonic
enhancement exists not only for the exciton $i=0$, but
also for any exciton which can be transformed into a $0$
exciton by Pauli scatterings with the sea excitons.

From a mathematical point of view, this extra $N$ is linked to
the topology of the diagrams representing $A_N(m,i)$. As in the
case of the well known Feynman diagrams for which superextensive
terms are linked to disconnected diagrams, we here see that an
extra factor $N$ appears in the part of $A_N(m,i)$ corresponding
to diagrams which are made of two pieces.

To conclude, we can say that the procedure we have used to
calculate $A_N(m,i)$ led us to represent this scalar product by
Pauli diagrams which are actually quite simple: The part of 
$A_N(m,i)$ which exists for any $\v Q_m=\v Q_i$ is made of all
connected diagrams with $m$ at the left bottom and $i$ at all
possible places on the right, the exciton lines being connected
by Pauli scatterings put in zigzag right, left, right\ldots (see
fig.\ 5c). $A_N(m,i)$ has an additional part when the $m$ and
$i$ excitons have a momentum equal to the sea exciton momentum
$\v Q_0$. This additional part is made of all possible Pauli
diagrams which can be cut into two pieces, $m$ staying at the
left bottom of one piece, while $i$ stays at the right bottom of
the other piece, the exciton lines being connected by Pauli
scatterings in zigzag right, left, right\ldots for the $m$
piece, and left, right, left\ldots for the $i$ piece (see fig.\
6c). As a direct consequence of the topology of these diconnected
diagrams, an extra factor $N$ appears in this part of
$A_N(m,i)$. This factor $N$ is physically linked to the well
known bosonic enhancement which, for composite excitons, exists
not only for an exciton identical to a sea exciton, but also for
any exciton which can be transformed into a sea exciton by Pauli
scatterings with the sea.

Although this result for the scalar product of $(N+1)$-exciton
states, with $N$ of them in the same state $0$, is nicely simple
at any order in Pauli interaction, it does not leave us
completely happy. Indeed, while in the diagrams which exist for
$\v Q_i=\v Q_m=\v Q_0$, the $m$ and $i$ indices play similar
roles, their roles in the diagrams which exist even if $\v
Q_i\neq \v Q_0$ are dissymmetric, which is not at
all satisfactory. This dissymmetry can be traced back to the way
we calculated 
$A_N(m,i)$. It is clear that equivalences between Pauli diagrams
have to exist in order to restore the intrinsic $(m,i)$ symmetry
of $A_N(m,i)$. In the last part of this work, we are going to
discuss some of these equivalences between Pauli diagrams.

However, the reader, not as picky as us, may just drop this last
part since, after all, the quite simple, although dissymmetric,
Pauli diagrams obtained above are enough to get the correct
answer for 
$A_N(m,i)$ at any order in the Pauli interactions.

\section{Equivalence between Pauli diagrams}

In order to have some ideas about which kinds of Pauli diagrams
can be equivalent, let us first derive the other possible
diagrammatic representations of $A_N(m,i)$. They use the
recursion relations between $A_N(m,i)$ and $A_{N-2}(m,j)$ or
$A_{N-2}(n,j)$, instead of $A_{N-2}(n,i)$.

\subsection{Pauli diagrams for $A_N(m,i)$ using $A_{N-2}(m,j)$}

To get this recursion relation, we must keep $B_m$ in the
calculation of $\hat{A}_N(m,i)$ defined in eq.\ (33). This leads
us to use $[B_0^N,B_i^\dag]$ instead of $[B_m,B_0^{\dag N}]$;
equation (38) is then replaced by
\begin{equation}
\hat{A}_N(m,i)=N\,c_N(m,i)+\frac{F_{N-2}}{F_N}\,N(N-1)\,\sum_{nj}
\lambda_{nj00}\,\lambda_{00in}\,A_{N-2}(m,j)\ ,
\end{equation}
in which we have set
\begin{equation}
c_N(m,i)=\frac{F_{N-1}}{F_N}\,\delta_{0i}\,\zeta_{N-1}(m)-
\frac{F_{N-2}}{F_N}\,(N-1)\lambda_{000i}\,\zeta_{N-2}(m)\ .
\end{equation}
Using fig.\ 2b for $\zeta_N(m)$, we easily obtain the diagrams
for $c_N(m,i)$ shown in fig.\ 7. When compared to
$b_N(m,i)$, we see that the roles played by $m$ and $i$ are
exchanged as well as the relative position of the crosses.

Equation (48) leads to write $A_N(m,i)$ as
\begin{equation}
A_N(m,i)=\overline{\alpha}_N(m,i)+N\,\overline{\beta}_N(m,i)\ ,
\end{equation}
where $\overline{\alpha}_N(m,i)$ and $\overline{\beta}_N(m,i)$
obey the recursion relations
\begin{equation}
\overline{\alpha}_N(m,i)=a_N(m,i)+\frac{F_{N-2}}{F_N}\,N(N-1)\,
\sum_{nj}\lambda_{nj00}\,\lambda_{00in}\,
\overline{\alpha}_{N-2}(m,j)\ ,
\end{equation}
\begin{equation}
\overline{\beta}_N(m,i)=c_N(m,i)+
\frac{F_{N-2}}{F_N}\,(N-1)(N-2)\,
\sum_{nj}\lambda_{nj00}\,\lambda_{00in}\,
\overline{\beta}_{N-2}(m,j)\ .
\end{equation}

Let us first consider $\overline{\beta}_N(m,i)$. As
$\beta_N(m,i)$, it differs from zero for $\v Q_m=\v Q_i=\v Q_0$
only. Its recursion relation leads to expand it in terms of
$c$'s as shown in fig.\ 8. If we now replace the $c$'s by
their expansion shown in fig.\ 7, we immediately find that 
$\overline{\beta}_N(m,i)$ is represented by the \emph{same}
Pauli diagrams as the ones for $\beta_N(m,i)$, so that 
$\overline{\beta}_N(m,i)=\beta_N(m,i)$. This is after all not
surprising because, in them, the roles played by $m$ and $i$ are
symmetrical.

From this result, we immediately conclude that the parts of
$A_N(m,i)$ which exist even if $\v Q_i=\v Q_m\neq \v Q_0$ have
also to be equal, \emph{i}.\ \emph{e}., we
must have $\overline{\alpha}_N (m,i)=\alpha_N(m,i)$. Let us now
see how this $\overline{\alpha} _N(m,i)$ appears, using eq.\
(51).

If we calculate $a_N(m,i)$ not with $[D_{mi},B_0^{\dag N}]$ but
with $[B_0^N,D_{mi}]$, we find that $a_N(m,i)$ can be
represented, not only by the diagrams of fig.\ 3b, but also
by those of fig.\ 3c. These diagrams look very similar,
except that the crosses are now in zigzag left, right,
left\ldots Since these two sets of diagrams (3b) and
(3c) represent the same
$a_N(m,i)$, while they have to be valid for $N=2,3,\cdots$,
the relative positions of the crosses have to be unimportant in
these Pauli diagrams. We will come back to this equivalence at
the end of this part.

The iteration of the recursion relation for $\overline{\alpha}_N
(m,i)$ leads to the diagrams of fig.\ 9a. If in them, we insert
the diagrams of fig.\ 3c for $a_N(m,i)$, we get the diagrams
of fig.\ 9b. They look like the ones for $\alpha_N(m,i)$,
except that $i$ now stays at the right bottom while $m$ moves
to all possible positions on the left, the zigzag for the
Pauli scatterings being now left, right, left\ldots This leads
to write $\overline{\alpha}_N(m,i)$ as
\begin{equation}
\overline{\alpha}_N(m,i)=\sum_{p=0}^N(-1)^p\,\frac{F_{N-p}}{F_N}
\,\frac{N!}{(N-p)!}\,\overline{Z}^{(p)}(m,i)\ ,
\end{equation}
where $\overline{Z}^{(p)}(m,i)$ represents the set of zigzag
diagrams of fig.\ 9b, with $p$ crosses.

Since $\overline{\alpha}_N(m,i)=\alpha_N(m,i)$, we conclude from
the validity of their expansions for $N=2,3,\cdots$, that the
zigzag diagrams $Z^{(p)}(m,i)$ and $\overline{Z}^{(p)}(m,i)$
must correspond to identical quantities, which is not obvious at
first.

\subsection{Pauli diagrams for $A_N(m,i)$ using $A_{N-2}(n,j)$}

To get this recursion relation, we start as for the one
between $A_N(m,i)$ and $A_{N-2}(n,i)$, but we use $[B_0^N,B_i
^\dag]$ instead of $[B_0^N,B_j^\dag]$ to calculate the matrix
element appearing in eq.\ (37). This leads to
\begin{equation}
\hat{A}_N(m,i)=N\,d_N(m,i)+\frac{F_{N-2}}{F_N}\,N(N-1)\,\sum_{nj}
\lambda_{mj00}\,\lambda_{00ni}\,A_{N-2}(n,j)\ ,
\end{equation}
in which we have set 
\begin{equation}
d_N(m,i)=\frac{F_{N-1}}{F_N}\,\delta_{m0}\,\zeta_{N-1}^\ast(i)
-\frac{F_{N-2}}{F_N}\,(N-1)\,\delta_{0i}\,\sum_j\lambda_{mj00}\,
\zeta_{N-2}^\ast(j)\ .
\end{equation}
Using the diagrams of fig.\ 2c for $\zeta_N^\ast(i)$, it is
easy to show that $d_N(m,i)$ is represented by the diagrams of
fig.\ 10. Note that they are different from the ones
for $b_N(m,i)$ and $c_N(m,i)$ shown in fig.\ 4b and fig.\ 7.
This is actually normal because, as seen from eq.\ (55), 
$d_N(m,i)$ is equal to zero when both $m\neq 0$ and $i\neq 0$,
while $b_N(m,i)$ and $c_N(m,i)$ differ from zero provided that
$\v Q_m (=\v Q_i)$ is equal to $\v Q_0$.

Equation (54) leads to write $A_N(m,i)$ as
\begin{equation}
A_N(m,i)=\overline{\overline{\alpha}}_N(m,i)+N\,
\overline{\overline{\beta}}_N(m,i)\ ,
\end{equation}
where $\overline{\overline{\alpha}}_N(m,i)$ and
$\overline{\overline{\beta}}_N(m,i)$ now obey
\begin{equation}
\overline{\overline{\alpha}}_N(m,i)=a_N(m,i)+\frac{F_{N-2}}{F_N}\,
N(N-1)\,\sum_{nj}\lambda_{mj00}\,\lambda_{00ni}\,
\overline{\overline{\alpha}}_{N-2}(n,j)\ ,
\end{equation}
\begin{equation}
\overline{\overline{\beta}}_N(m,i)=d_N(m,i)+\frac{F_{N-2}}{F_N}\,
(N-1)(N-2)\,\sum_{nj}\lambda_{mj00}\,\lambda_{00ni}\,
\overline{\overline{\beta}}_{N-2}(n,j)\ .
\end{equation}

Let us start with $\overline{\overline{\beta}}_N(m,i)$. Its
recursion relation is shown in fig.\ 11a, as well as its
iteration (fig.\ 11b). (In it, we have used the two
equivalent forms of this recursion relation given
in fig.\ 11a). If, in these diagrams, we now insert the
diagrammatic representation of $d_N(m,i)$ shown in fig.\ 10,
with the $m$ part alternatively below and above the $i$ part, we
find that 
$\overline{\overline{\beta}}_N(m,i)$ is represented by exactly
the same diagrams as the ones for $\beta_N(m,i)$, so that 
$\overline{\overline{\beta}}_N(m,i)=\beta_N(m,i)$. As a
consequence, we must have 
$\overline{\overline{\alpha}}_N(m,i)=\alpha_N(m,i)$.

Let us now consider the recursion relation between 
$\overline{\overline{\alpha}}_N(m,i)$ and 
$\overline{\overline{\alpha}}_{N-2}(n,j)$. The iteration of this
recursion relation leads to the diagrams of fig.\ 12a. If in
it, we insert the diagrammatic representation of $a_N(m,i)$ shown
in fig.\ 3b, we get the diagrams of fig.\ 12b. We can note
that it is not enough to use $\lambda_{0n00}=\lambda_{n000}$ to
transform the last third order diagram into
the two last third order zigzag diagrams left, right, left of
$\alpha_N(m,i)$. At fourth order, the situation is even worse,
the last fourth order Pauli diagram for
$\overline{\overline{\alpha}}_N(m,i)$ being totally different
from a zigzag diagram. They however have to represent the same
quantity because
$\overline{\overline{\alpha}}_N(m,i)=\alpha_N(m,i)$ for any $N$.

Let us now identify the underlying reason for the equivalence of
Pauli diagrams like the ones of figs.\ 3b and 3c which
represent the same $a_N(m,i)$, or the ones of figs.\ 5c, 9b
and 12b which represent the part of the same $A_N(m,i)$ which
exists even if $\v Q_m(=\v Q_i)\neq \v Q_0$. This will help us to
understand how these Pauli diagrams really work.

\subsection{``Exchange skeletons''}

All the Pauli diagrams we found in the preceding sections, are
made of a certain number of exciton lines connected by Pauli
scatterings between two excitons, put in various orders. It is
clear that the value of these diagrams can depend on the ``in''
and ``out'' exciton states, \emph{i}.\ \emph{e}., the indices
which appear at the right and the left of these diagrams, but
not on the intermediate exciton states over which sums are
taken. Between these ``in'' and ``out'' excitons, a lot of
carrier exchanges take place through the various Pauli
scatterings represented by crosses in the Pauli diagrams. It is
actually reasonable to think that these Pauli diagrams have to
ultimately read in terms of ``exchange skeletons'' between the
``in'' and ``out'' excitons of these diagrams. For $N=2, 3,
4,$\ldots excitons, these ``exchange skeletons'' should appear as
\begin{equation}
L^{(2)}(m_1,m_2;i_1,i_2)=\int d\v r_{e_1}\cdots d\v
r_{h_2}\phi_{m_1}^\ast(\v r_{e_1},\v r_{h_1})
\phi_{m_2}^\ast(\v r_{e_2},\v r_{h_2})
\phi_{i_1}(\v r_{e_1},\v r_{h_2})
\phi_{i_2}(\v r_{e_2},\v r_{h_1}),
\end{equation}
\begin{eqnarray}
L^{(3)}(m_1,m_2,m_3;i_1,i_2,i_3)=\int d\v r_{e_1}\cdots d\v
r_{h_3}
\phi_{m_1}^\ast(\v
r_{e_1},\v r_{h_1})\phi_{m_2}^\ast(\v r_{e_2},\v r_{h_2})
\phi_{m_3}^\ast(\v r_{e_3},\v r_{h_3})\nonumber \\ \times
\phi_{i_1}(\v r_{e_1},\v r_{h_2})
\phi_{i_2}(\v r_{e_3},\v r_{h_1})\phi_{i_3}(\v r_{e_2},\v
r_{h_3}),
\end{eqnarray}
\begin{eqnarray}
L^{(4)}(m_1,m_2,m_3,m_4;i_1,i_2,i_3,i_4)=\int d\v
r_{e_1}\cdots d\v r_{h_4}
\phi_{m_1}^\ast(\v
r_{e_1},\v r_{h_1})
\phi_{m_2}^\ast(\v r_{e_2},\v r_{h_2})\hspace{2cm}\nonumber \\
\times
\phi_{m_3}^\ast(\v r_{e_3},\v r_{h_3})
\phi_{m_4}^\ast(\v r_{e_4},\v r_{h_4})
\phi_{i_1}(\v r_{e_1},\v r_{h_2})
\phi_{i_2}(\v r_{e_3},\v r_{h_1})\phi_{i_3}(\v r_{e_2},\v
r_{h_4})\phi_{i_4}(\v r_{e_4},\v r_{h_3}),
\end{eqnarray}
and so on\ldots These definitions are actually  transparent
once we look at the diagrammatic representations of these
exchange skeletons shown in fig.\ 13.

There are in fact various equivalent ways to represent these
exchange skeletons, as can be seen from fig.\ 14 in the case
of three excitons. These equivalent representations simply say
that the
$i_1$ exciton has the same electron as the $m_1$ exciton and the
same hole as the $m_2$ exciton, so that $i_1$ and $m_1$ must be
connected by an  electron line while $i_1$ and $m_2$ must be
connected by a hole line.

All possible carrier exchanges between $N$ excitons can be
expressed in terms of these exchange skeletons. 

\begin{itemize} 
\item In the
case of two excitons, the Pauli scattering which
appears in the Pauli diagrams, is just
$\lambda_{mnij}=\left(L^{(2)}(m,n;i,j)+L^{(2)}(n,m;i,j)
\right)/2$ (see fig.\ 1a). In our many-body theory for
interacting composite bosons, this $\lambda_{mnij}$ appears
as composed of processes in which the
indices
$m$ and $n$ are exchanged. This is actually equivalent to say
that the excitons exchange their electrons instead of their
holes (see fig.\ 1b). We can also note that, when the two
indices on one side are equal, like in
$\lambda_{mn00}$, to exchange an electron or to exchange a hole
is just the same (see figs.\ 1d and 1e).

\item For three excitons, we could think of a carrier
exchange between the $(i,j,k)$ and $(m,n,p)$ excitons, different
from the one corresponding to $L^{(3)}(m,n,p;i,j,k)$. Let us,
for example, consider the one which would read like eq.\ (60),
with
$(m_1,m_2,m_3,i_1)$ respectively replaced by $(m,n,p,i)$, while
$\phi_{i_2}(\v r_{e_3},\v r_{h_1})\phi_{i_3}(\v r_{e_2},\v
r_{h_3})$ is replaced by $\phi _j(\v r_{e_2},\v
r_{h_3})\phi_k(\v r_{e_3},\v r_{h_1})$. This carrier exchange,
shown in fig.\ 15, indeed reads as an exchange skeleton, being
simply $L^{(3)}(m,n,p;i,k,j)$.

And so on, for any other carrier exchange we could think of.
\end{itemize}

Let us now consider the various diagrams we have found in
calculating $A_N(m,i)$ and understand why they are
indeed equivalent, in the light of these exchange skeletons.

\subsection{Pauli diagrams with one exciton only different from
0}

We first take the simplest of these Pauli diagrams, namely
the zigzag diagram $z^{(p)}(m,0)$ entering $\zeta_N(m)$, shown
in fig.\ 2. On its left, this zigzag diagram has $p$ excitons $0$
and one exciton $m$, while on its right, the $(p+1)$ excitons
are all $0$ excitons. After summation over the intermediate
exciton indices, the final expression of this Pauli diagram must
read as an integral of $\phi_m^\ast(\v r_{e_1},\v
r_{h_1})\phi_0^\ast (\v r_{e_2},\v r_{h_2})$ \ldots
$\phi_0^\ast(\v r_{e_{p+1}},
\v r_{h_{p+1}})$ multiplied by $(p+1)$ wave functions $\phi_0$
with the $(\v r_{e_i},\v r_{h_i})$'s mixed in such a way that
the integral cannot be cut into two independent integrals
(otherwise the Pauli diagram would be topologically
disconnected). This is exactly what the exchange skeleton
$L^{(p+1)}(m,0\cdots,0;0,\cdots,0)$ does.The possible
permutations of the various $(\v r_e,\v r_h)$'s in the definition
of this 
$L^{(p+1)}$, actually show that the $m$ index in
the diagrammatic representation of this exchange skeleton can be
in any possible place on the left. This is
also true for the $m$ position in the Pauli diagrams with all the
indices equal to
$0$ except one. Moreover, the relative position of the
crosses in these diagrams are unimportant (see fig.\ 16). This
is easy to show, just by ``sliding'' the carrier
exchanges, as explicitly shown in the case of three excitons
(see fig.\ 17). In this figure, we have also used the fact that
the Pauli scatterings reduce to one diagram when the two indices
on one side are equal (see figs.\ 1d and 1e). This possibility
to ``slide'' the carrier exchange, mathematically comes from the
fact that $\sum_q\phi_q^\ast(\v r_e,\v r_h)\phi_q(\v r_{e'},
\v r_{h'})$ is nothing but $\delta(\v r_e-\v r_{e'})\delta
(\v r_h-\v r_{h'})$.

\subsection{Pauli diagrams with one exciton on each side
different from 0}

There are essentially two kinds of such diagrams: Either the two
excitons $(m,i)$ different from $0$ have no common carrier, or
they have one. Let us start with this second case.

\subsubsection{$m$ and $i$ have one common carrier}

Two different exchange skeletons exist in this case, depending
if the common carrier is an electron or a hole. They are shown
in figs.\ 18a and 18b. By ``sliding'' the carrier exchanges
as done in fig.\ 18c, it is easy to identify the set of Pauli
diagrams which correspond to the sum of these two exchange
skeletons (see fig.\ 18d). This in particular shows the
identity of the Pauli diagrams of fig.\ 18e which enter the
two diagrammatic representations of $a_N(m,i)$ shown in figs.\
3b and 3c.

\subsubsection{$m$ and $i$ do not have a common carrier}

In this case, the number of different exchange skeletons depends
on the number of $0$ excitons involved.

\begin{itemize}

\item For two $0$ excitons, there is one exchange skeleton only
(see fig.\ 19). By ``sliding'' the carrier exchanges, we get
the two equivalent Pauli diagrams shown in fig.\ 19. They can
actually be deduced from one another just by symmetry up/down,
which results from $\lambda_{mnij}=\lambda_{nmji}$.

\item For three $0$ excitons, there are two possible exchange
skeletons which are actually related by electron-hole
exchange (see figs.\ 20a and 20b). By sliding the carrier
exchanges, we get the two diagrams of fig.\ 20c, so that, by
combining these two exchange skeletons, we find the two Pauli
diagrams of fig.\ 20d.

\item In the same way, for four $0$ excitons, there are three
possible exchange skeletons, two of them being related by
electron-hole exchange (see figs.\ 21a,21b,21c). By sliding the
carrier exchanges, it is again possible to find the Pauli
diagrams corresponding to these exchange skeletons as shown in
fig.\ 21.

\end{itemize}

\subsection{Equivalent representations of the diagrams appearing
in $A_N(m,i)$}

By expressing the various Pauli diagrams entering the expansion
of $A_N(m,i)$ in terms of these exchange skeletons, as shown in
figs.\ (16-21), it is now possible to directly prove their
equivalence.

Fig. 22 shows the set of transformations which allows to go
from the last third order diagrams of
$\overline{\overline{\alpha}}_N(m,i)$ to the zigzag Pauli
diagrams of $\alpha_N(m,i)$, with $i$ at the two upper positions.

The transformation of the last fourth order diagram
for $\overline{\overline{\alpha}}_N(m,i)$ into the two missing
zigzag diagrams of $\alpha_N(m,i)$ is somewhat more subtle. For
the interested reader, let us describe in details how this can be
done. We have reproduced in fig.\ 23, the last fourth order
diagram of
$\overline{\overline{\alpha}}_N(m,i)$, as it
appears in fig.\ 12b. We see that all Pauli
scatterings have two $0$ indices on one side, except
$\lambda_{npq0}$. According to figs.\ 1a,1b, this
$\lambda_{npq0}$ can be represented by the sum of an
electron exchange plus a hole exchange between the $(q,0)$
excitons. By continuity, we represent each of the other Pauli
scatterings which have two $0$ excitons on one side, either by
fig.\ 1d or by fig.\ 1e, the choice between these two
equivalent representations being driven by avoiding the crossings
of electron and hole lines. This leads to the two diagrams
of figs.\ 23b and 23c: They just correspond to 
exchange the role played by the electrons and the holes.

We now consider one of these  two diagrams, namely the one of
fig.\ 23b. Let us call
$0$,
$0'$, $0''$ and $0'''$ the four \emph{identical} $0$ excitons on
the right, in order to recognize them more easily when we will
redraw this diagram. To help visualizing this
redrawing, we have also given names to the various carriers. It
is then straightforward to check that the diagram of fig.\ 23b
corresponds to the samecarrier exchanges as the ones of
fig.\ 23d. Since the diagram of fig.\ 23c is the same as the
one of fig.\ 23b, with the electrons replaced by holes, we
conclude that twice the ugly diagram of fig.\ 23a is indeed equal
to the two missing zigzag diagrams of
$\alpha_N(m,i)$ reproduced in the top of this figure.

\section{Conclusion}

In this paper, we have essentially calculated the scalar product
of $(N+1)$-exciton states with $N$ of them in the same state
$0$. This scalar product is far from trivial due to many-body
effects induced by ``Pauli scatterings'' which originate from
the composite nature of excitons. As a result, these scalar
products appear as expansions in
$\eta=N\mathcal{V}_X/\mathcal{V}$, where $\mathcal{V}_X$ and
$\mathcal{V}$ are the exciton and sample volumes, with
possibly some additional factors $N$.

In order to understand the physical origin of these additional
$N$'s --- which will ultimately differentiate
superextensive from regular terms --- we have introduced the
concept of exciton dressed by a sea of excitons,
$$|\psi_i^{(N)}\rangle=\frac{B_0^NB_0^{\dag N}}{\langle v|B_0^N
B_0^{\dag N}|v\rangle}\ B_i^\dag|v\rangle\ .$$
In the absence of Pauli interaction between the exciton $i$ and
the sea of $N$ excitons $0$, the operator in front of $B_i^\dag$
reduces to an identity. Due to Pauli interaction, contributions
on other exciton states $B_{m\neq i}^\dag|v\rangle$ appear in
$|\psi_i^{(N)}\rangle$, which originate from possible carrier
exchanges between the $i$ exciton and the sea. We moreover find that a
bosonic enhancement, which gives rise to an extra factor $N$, --- reasonable 
for the exciton
$i=0$, since after all, excitons are not so far from
bosons --- also exists when the $i$ exciton can be
transformed by carrier exchanges into a sea exciton. This
happens for any exciton $i$ having the same
center of mass momentum as the sea exciton. 

In order to understand the carrier exchanges between $N$
excitons, which make the scalar products of $N$-exciton states so
tricky, we have introduced \emph{``Pauli diagrams''}. They read in
terms of \emph{Pauli scatterings between two excitons}.
With them, we have shown how to generate a diagrammatic
representation of the scalar products of $(N+1)$-exciton states
with $N$ of them in the same state $0$, at any order in Pauli
interaction.

This diagrammatic representation is actually not unique.
Although the one we first give, is nicely simple to
memorize, more complicated ones, obtained from other procedures
to calculate the same scalar product, are equally good in the
sense that they lead to the same correct result.

In order to understand the equivalence between these various
Pauli diagrams, we have introduced \emph{``exchange skeletons''}
which correspond to \emph{carrier exchanges between more than two
excitons}. Their appearance in the scalar products
of
$N$-exciton states is actually quite reasonable because, even if
we can calculate these scalar products in terms of Pauli
scatterings between \emph{two} excitons only, Pauli exclusion is
originally $N$-body ``at once'': When a new exciton is
added, its carriers must be in states different from the
ones of all the previous excitons. The Pauli scatterings
between two excitons generated by our many-body theory for
interacting composite bosons, are actually quite convenient 
to calculate many-body effects between excitons at any
order in the interactions. It is however reasonable to find that
a set of such Pauli scatterings, which in fact correspond to
carrier exchanges between more than two excitons, finally read
in terms of these ``exchange skeletons''.

The present work is restricted to scalar products of $N$-exciton
states in which all of them, except one, are in the same $0$ state.
In physical effects involving $N$ excitons, of course enter more complicated
scalar products. Such scalar products will be presented in a forthcoming
publication. The detailed study presented here,
is however all the more useful, because it
allows to identify the main characteristics of these scalar
products, which are actually present in more complicated situations: As
an example, in the case of two dressed excitons
$(i,j)$, a bosonic enhancement is found not only for $\v
Q_i$ or $\v Q_j$ equal to $\v Q_0$, but also for $\v Q_i+\v
Q_j=2\v Q_0$, because these
$(i,j)$ excitons can transform themselves into two $0$ excitons
by carrier exchanges. The corresponding processes are
represented by topologically disconnected Pauli diagrams, and extra
factors $N$ appear as a signature of this topology.

\newpage

\vspace{0.5cm}

\hbox to \hsize {\hfill REFERENCES
\hfill}

\noindent
[1] M. Combescot, C. Tanguy, \emph{Europhys.\ Lett.\ }\textbf{55}, 390 (2001).

\noindent
[2] M. Combescot, O. Betbeder-Matibet, \emph{Europhys.\ Lett.\ }\textbf{58},
87 (2002).

\noindent
[3] O. Betbeder-Matibet, M. Combescot, \emph{Eur.\ Phys.\ J.\ B} \textbf{27},
505 (2002).

\noindent
[4] M. Combescot, O. Betbeder-Matibet, \emph{Europhys.\ Lett.\ }\textbf{59},
579 (2002).

\noindent
[5] M. Combescot, X. Leyronas, C. Tanguy, \emph{Eur.\ Phys.\ J.\ B} \textbf{31},
17 (2003).

\noindent
[6] O. Betbeder-Matibet, M. Combescot, \emph{Eur.\ Phys.\ J.\ B} \textbf{31},
517 (2003).

\noindent
[7] M. Combescot, O. Betbeder-Matibet, K. Cho, H. Ajiki, Cond-mat/0311387.

\noindent
[8] A.A. Abrikosov, L.P. Gorkov, I.E. Dzyaloshinski, \textbf{Methods of quantum
field theory in statistical physics}, \emph{Prentice-hall, inc. Englewood
cliffs N.J.} (1964).

\noindent
[9] C. Cohen-Tannoudji, B. Diu, F. Lalo\"{e}, \textbf{M\'{e}canique Quantique},
\emph{Hermann, Paris} (1973).

\newpage

\begin{figure}
\centerline{ \scalebox{0.6}{\includegraphics{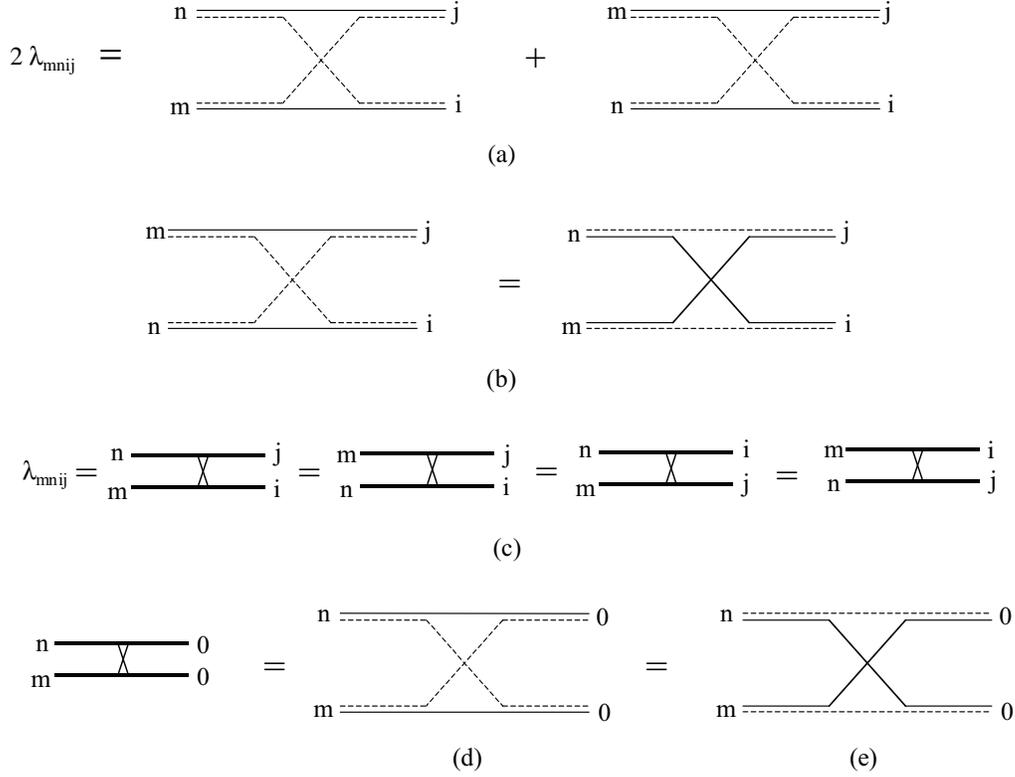}}}
\caption{\footnotesize{
Pauli scattering $\lambda_{mnij}$ between the ``in''
excitons
$(i,j)$ and the ``out'' excitons $(m,n)$. Solid line: electron; dashed
line: hole; heavy solid line: exciton. (a): As defined in eq.\ (3),
$\lambda_{mnij}$ is composed of two hole exchanges, with the
exciton indices
$(m,n)$ inverted, so that
$\lambda_{mnij}=\lambda_{nmij}=\lambda_{mnji}$.
(b): A hole exchange between $(i,j)$ and $(n,m)$
corresponds to an electron exchange between $(i,j)$ and $(m,n)$.
(c): In the Pauli diagrams of the following figures,
the Pauli scatterings $\lambda_{mnij}$ will be represented by
crosses.
(d) and (e): When the two indices on one side are
equal, the two processes of $\lambda_{mnij}$ are identical:
$\lambda_{mnij}$ either corresponds to a hole exchange as in (d) or an
electron exchange as in (e).}}
\end{figure}

\clearpage

\begin{figure}
\centerline{ \scalebox{0.7}{\includegraphics{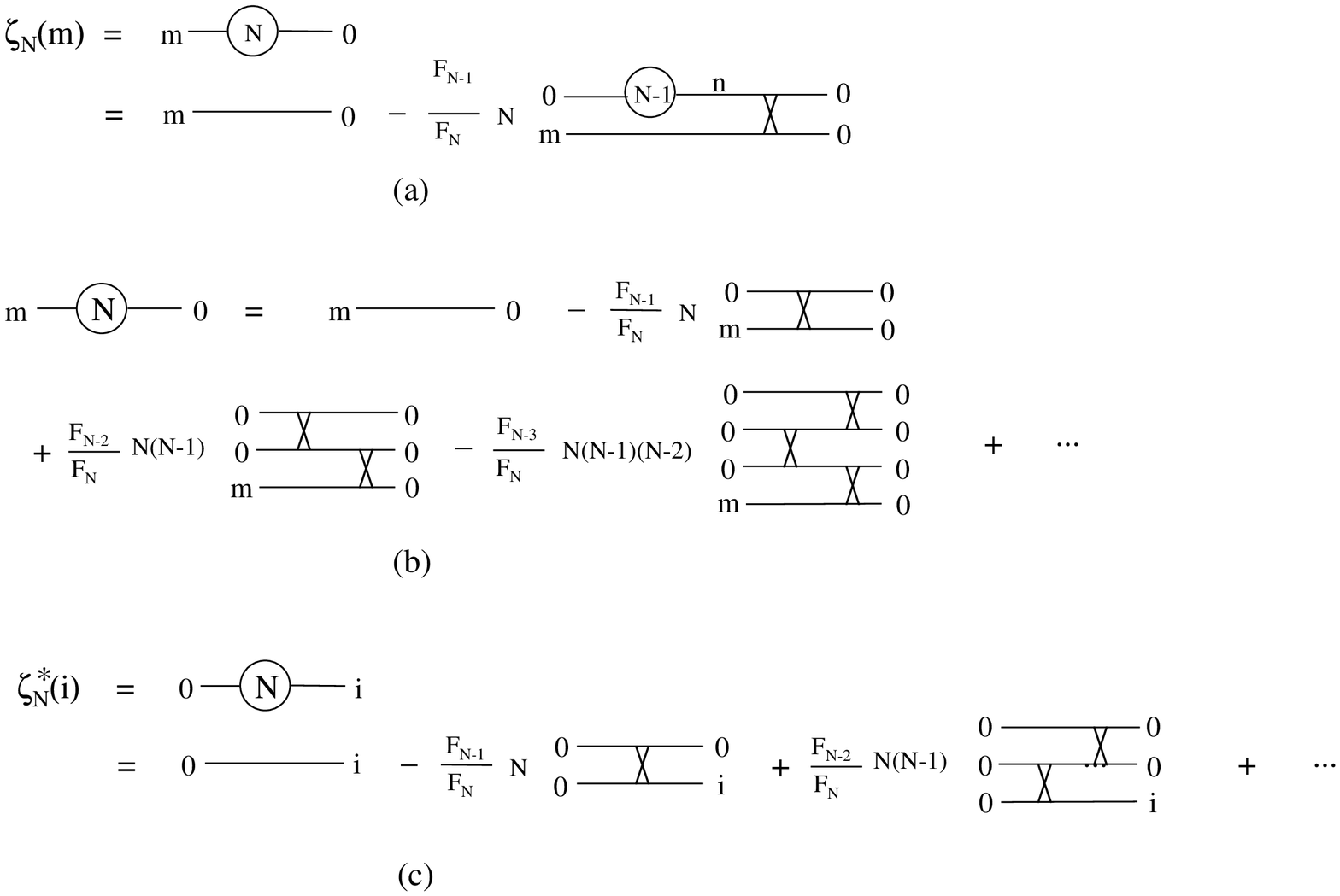}}}
\caption{\footnotesize{
(a): Recursion relation for the matrix element
$\zeta_N(m)$ defined in eq.\ (26). The $(m,0)$ line
represents $\delta_{m0}$, while the cross represents the Pauli
scattering $\lambda_{mn00}$ (see fig.\ 1). Sum is taken over the
intermediate $n$ exciton.
(b): Pauli diagrams for $\zeta_N(m)$ obtained by
iteration of fig.\ 2a. They allow to visualize eq.\ (28) in a
simple way. Sums are taken over the (unlabelled) exciton lines.
(c): The complex conjugate of $\zeta_N(i)$ is
represented by similar zigzag diagrams; the relative position
of the crosses are just changed, being alternatively left, right,
left,\ldots, instead of right, left, right,\ldots as in (b).
}}
\end{figure}

\clearpage

\begin{figure}
\centerline{ \scalebox{0.8}{\includegraphics{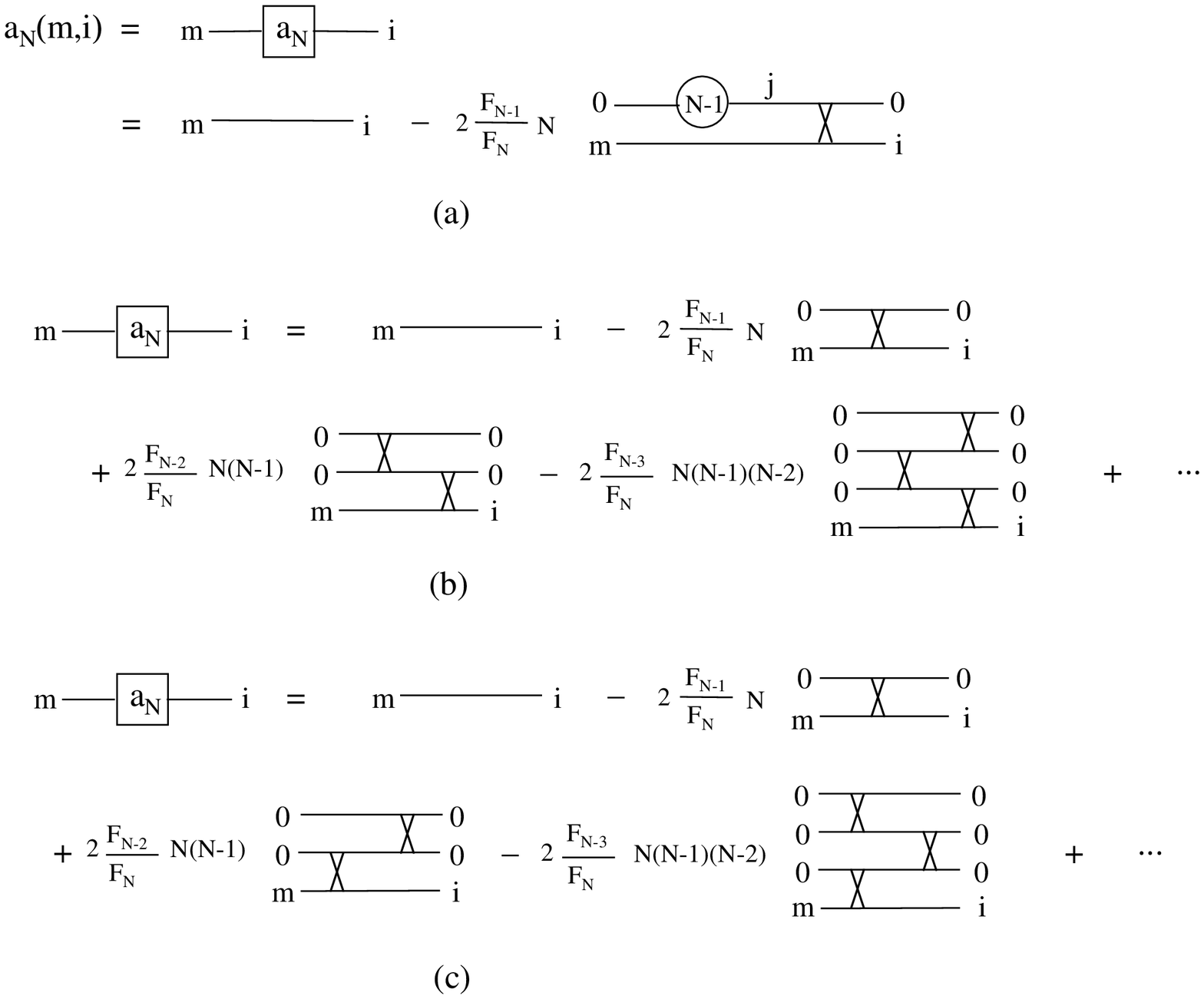}}}
\caption{\footnotesize{
(a): Diagrammatic representation of eq.\ (34), which gives the
quantity 
$a_N(m,i)$ defined in eq.\ (32).
(b): Pauli diagrams for this $a_N(m,i)$. They are obtained
by inserting diagrams of fig.\ 2c for $\zeta^\ast$ into fig.\ 3a.
They contain 0,1,2,3,\ldots Pauli scatterings represented by
crosses, put in zigzag, alternatively right, left, right,\ldots
.
$m$ and $i$ stay on the bottom line. Fig.\ 3b is just a 
visualization of eq.\ (35). Here again, as in all the Pauli
diagrams, sums are taken over the intermediate (unlabelled)
exciton lines.
(c): Same $a_N(m,i)$ obtained by using $[B_0^N,D_{mi}]$, instead
of $[D_{mi},B_0^{\dag N}]$, in eq.\ (32). This
other diagrammatic representation of $a_N(m,i)$, in which the
relative position of the crosses is just changed compared to
fig.\ 3b, is the one useful to get
$A_N(m,i)$ in terms of
$A_{N-2}(m,j)$, as done in the last part of this work.
}}
\end{figure}

\clearpage

\begin{figure}
\centerline{ \scalebox{0.7}{\includegraphics{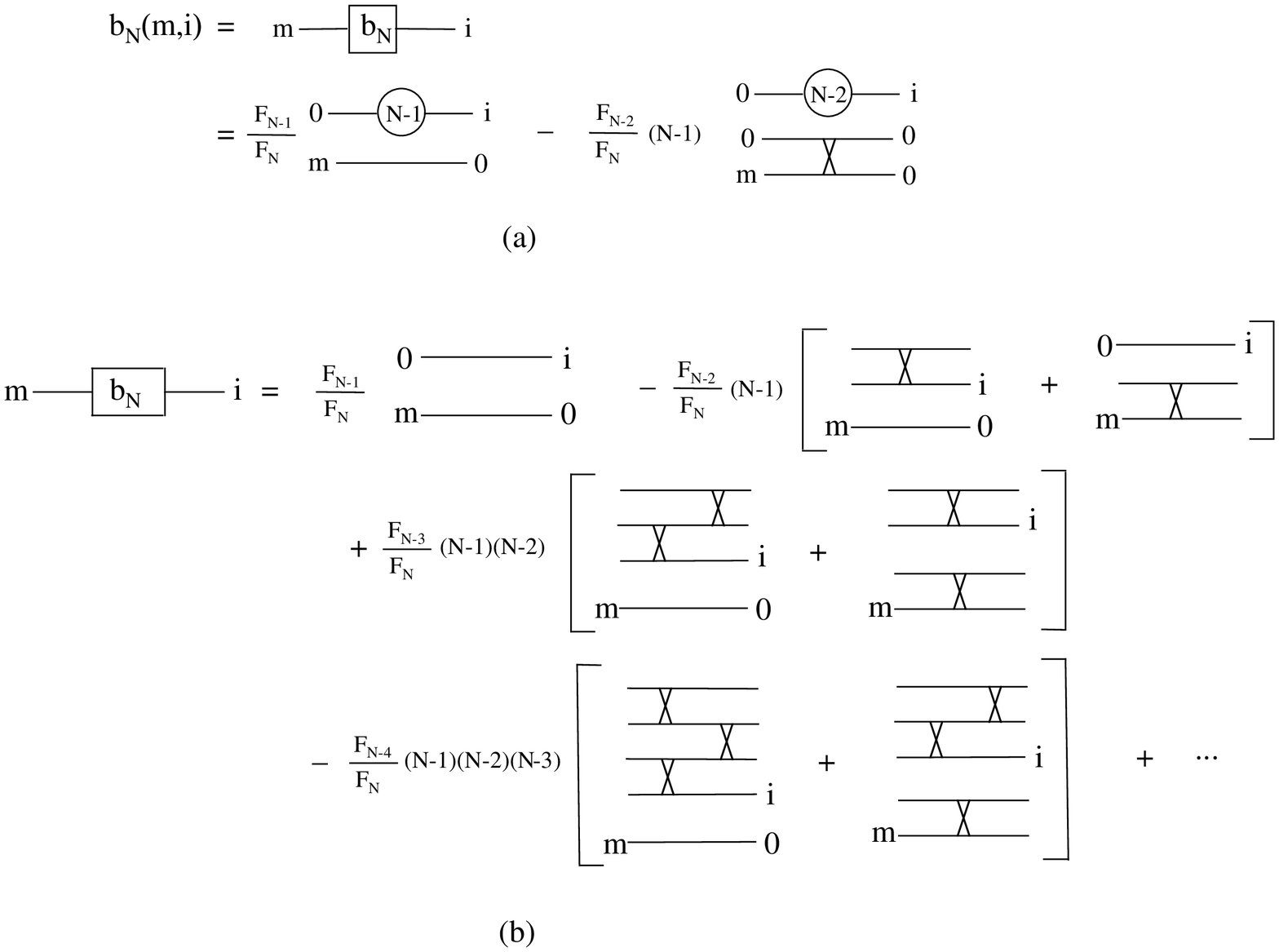}}}
\caption{\footnotesize{(a): Diagrammatic representation of eq.\ (39) which
defines the quantity $b_N(m,i)$.
(b): Pauli diagrams for this $b_N(m,i)$ obtained by
inserting the diagrams of fig.\ 2c for $\zeta^\ast$ into fig.\
4a. 
$b_N(m,i)$ is made of a set of \emph{disconnected} diagrams, with
0,1,2,3,\ldots Pauli scatterings represented by crosses. At
each order, it contains two terms, one with a $\delta_{m0}$
factor, the other with a $\lambda_{m000}$ factor. As in the
following Pauli diagrams, we have omitted the final exciton
indices $0$ to simplify the figures. Fig.\ 4b is a simple
visualization of eq.\ (40).}}
\end{figure}

\clearpage

\begin{figure}
\centerline{ \scalebox{0.55}{\includegraphics{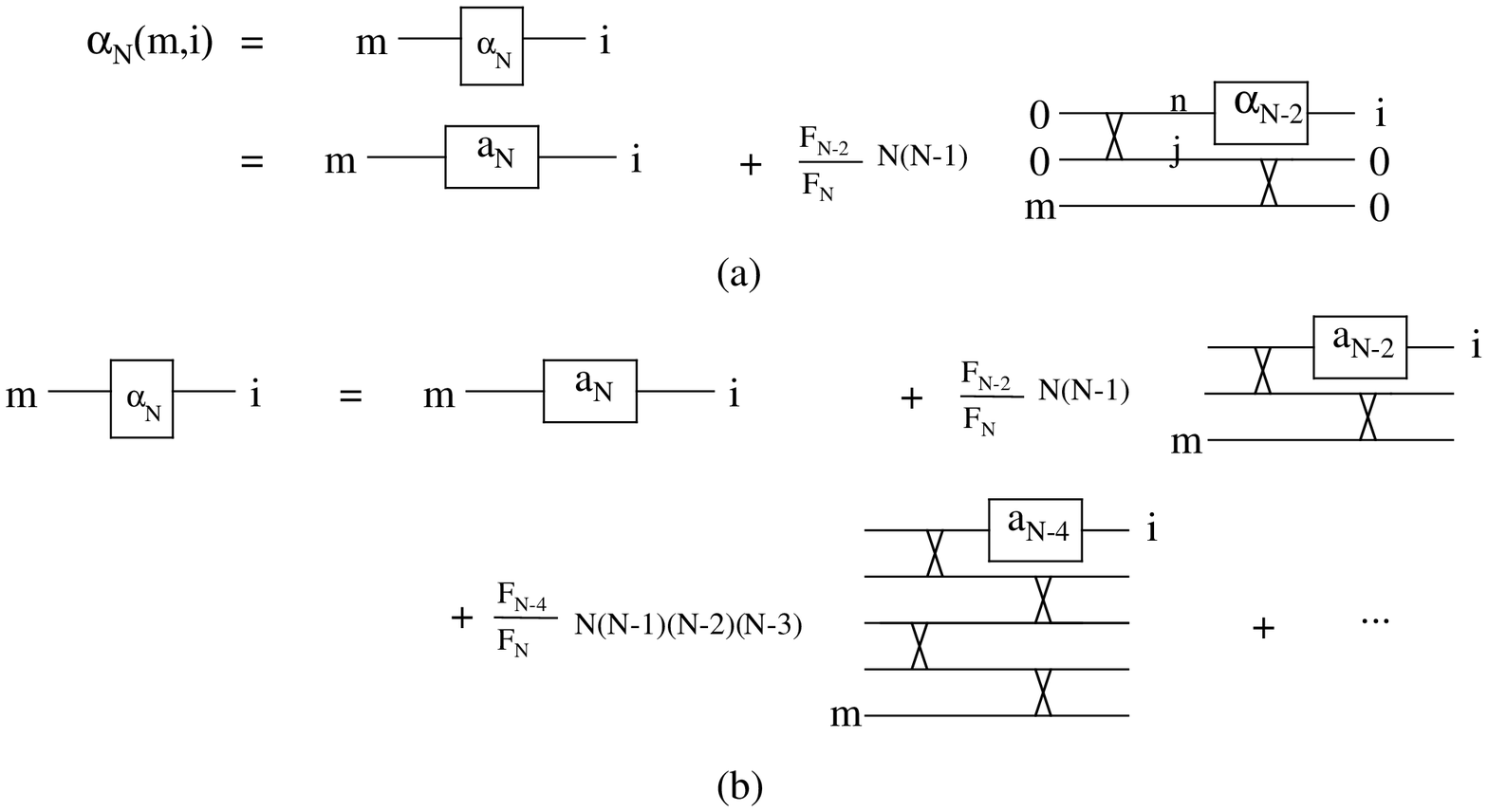}}}
\centerline{\scalebox{0.55}{\includegraphics{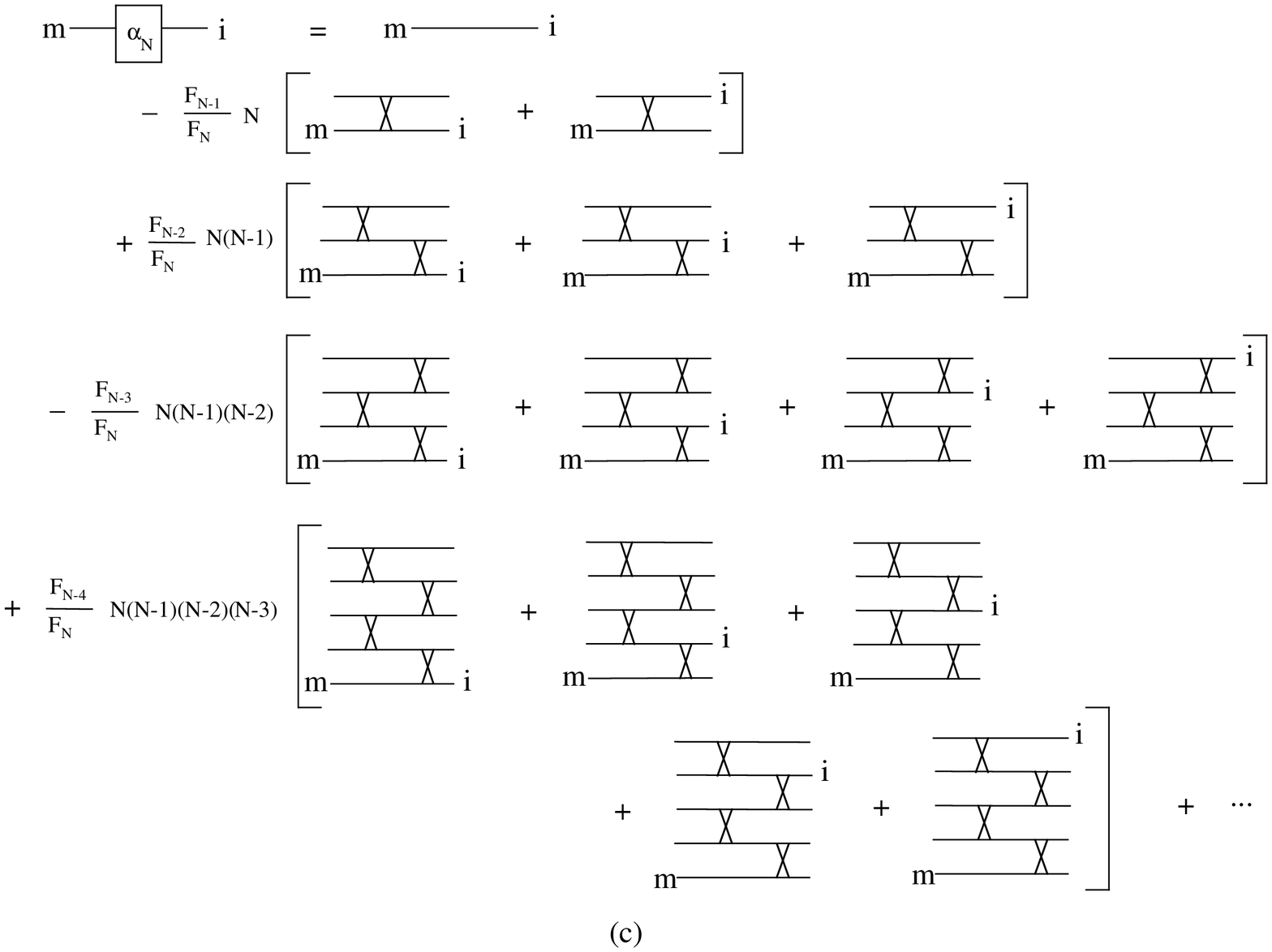}}}
\caption{\footnotesize{(a): Recursion relation (43) for the part
$\alpha_N(m,i)$ of the scalar product $A_N(m,i)$ which exists
even if $\v Q_m=\v Q_i\neq \v Q_0$.
(b): Iterative expansion of $\alpha_N(m,i)$ in terms of
$a_N(m,i)$.
(c):The Pauli diagrams for $\alpha_N(m,i)$, obtained by inserting
the diagrammatic representation of $a_N(m,i)$ shown in fig.\ 3b,
into fig.\ 5b, are made of zigzag diagrams with
Pauli scatterings put alternatively right, left, right,\ldots
The index $m$ stays at the left bottom, while $i$ moves at all
possible positions on the right. Fig.\ 5c is just a
visualization of eq.\ (45).}}
\end{figure}

\clearpage

\begin{figure}
\centerline{ \scalebox{0.55}{\includegraphics{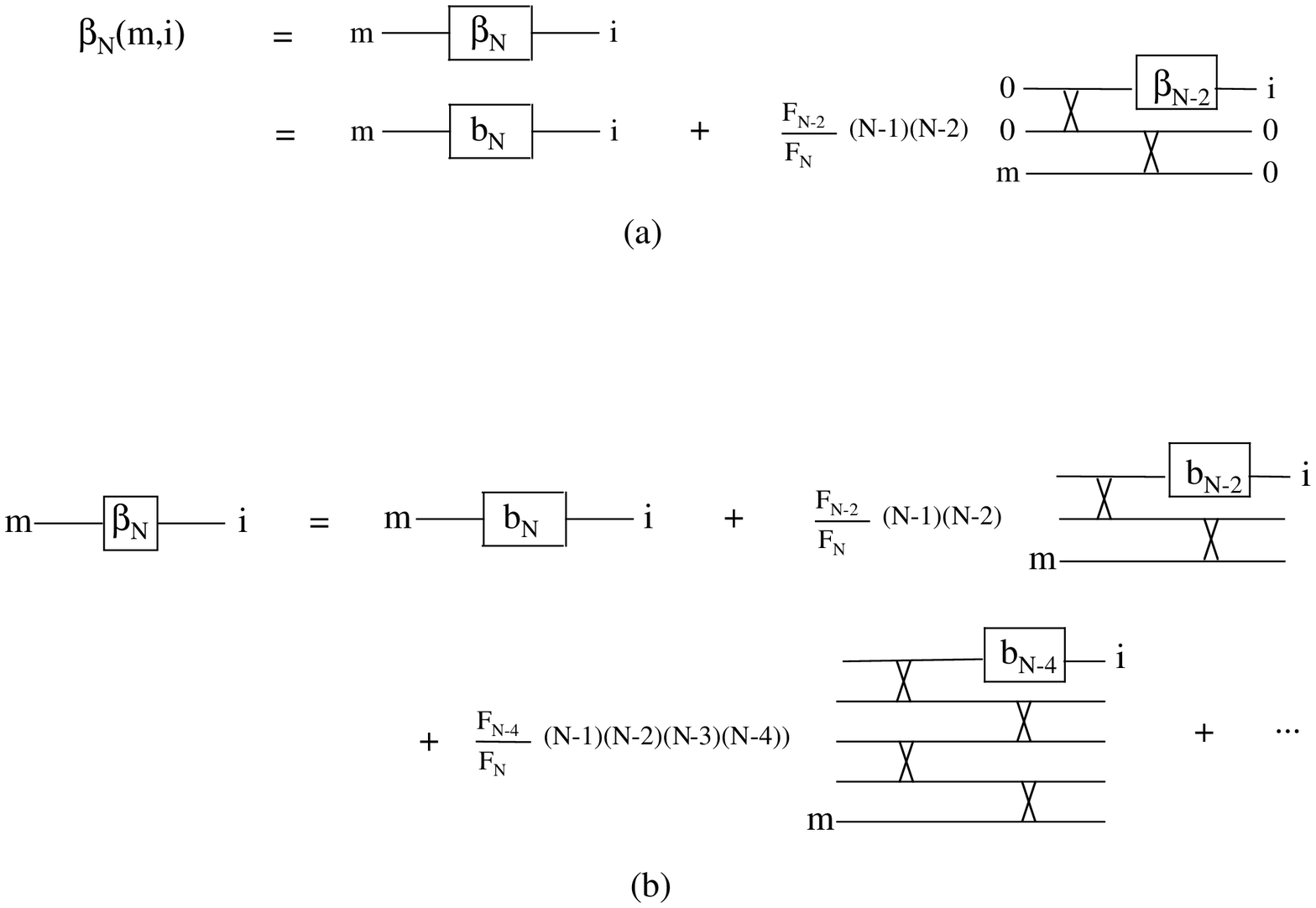}}}
\centerline{\scalebox{0.55}{\includegraphics{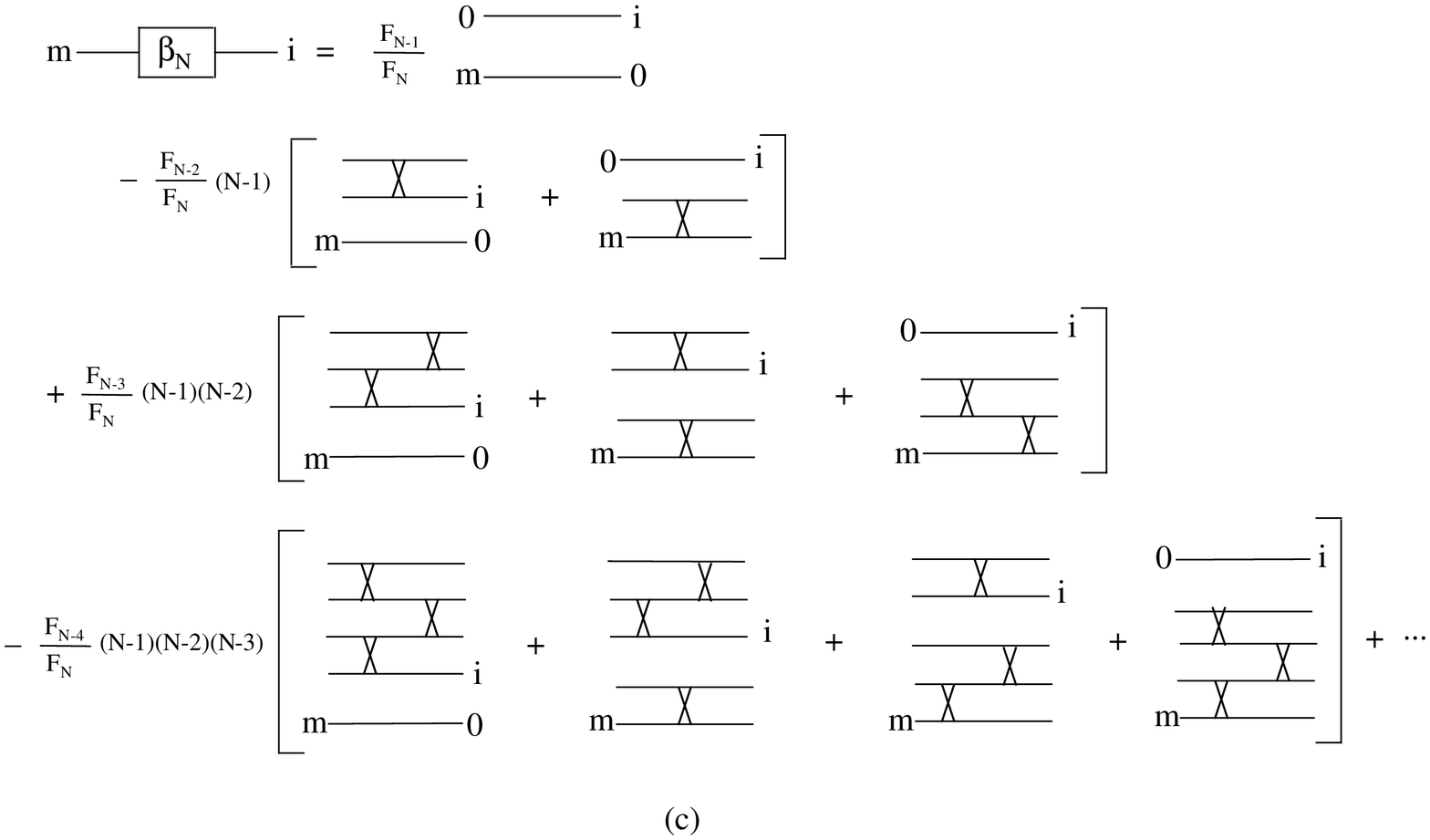}}}
\caption{\footnotesize{
(a): Recursion relation (44) for the part $\beta_N(m,i)$ of the
scalar product $A_N(m,i)$ which exists for $\v Q_m=\v Q_i=\v
Q_0$ only.
(b): Iterative expansion of $\beta_N(m,i)$ in terms of
$b_N(m,i)$.
(c): Pauli diagrams for $\beta_N(m,i)$ obtained by inserting the
diagrammatic representation of $b_N(m,i)$ shown in fig.\ 4b,
into fig.\ 6b. $\beta_N(m,i)$ is made of \emph{disconnected}
diagrams, the two pieces being zigzag diagrams right, left,
right,\ldots for the part with $m$ at the left bottom, and left,
right, left,\ldots for the part with $i$ at the right bottom.
Fig.\ 6c is just a visualization of eq.\ (47).
}}
\end{figure}

\clearpage

\begin{figure}
\centerline{\scalebox{0.7}{\includegraphics{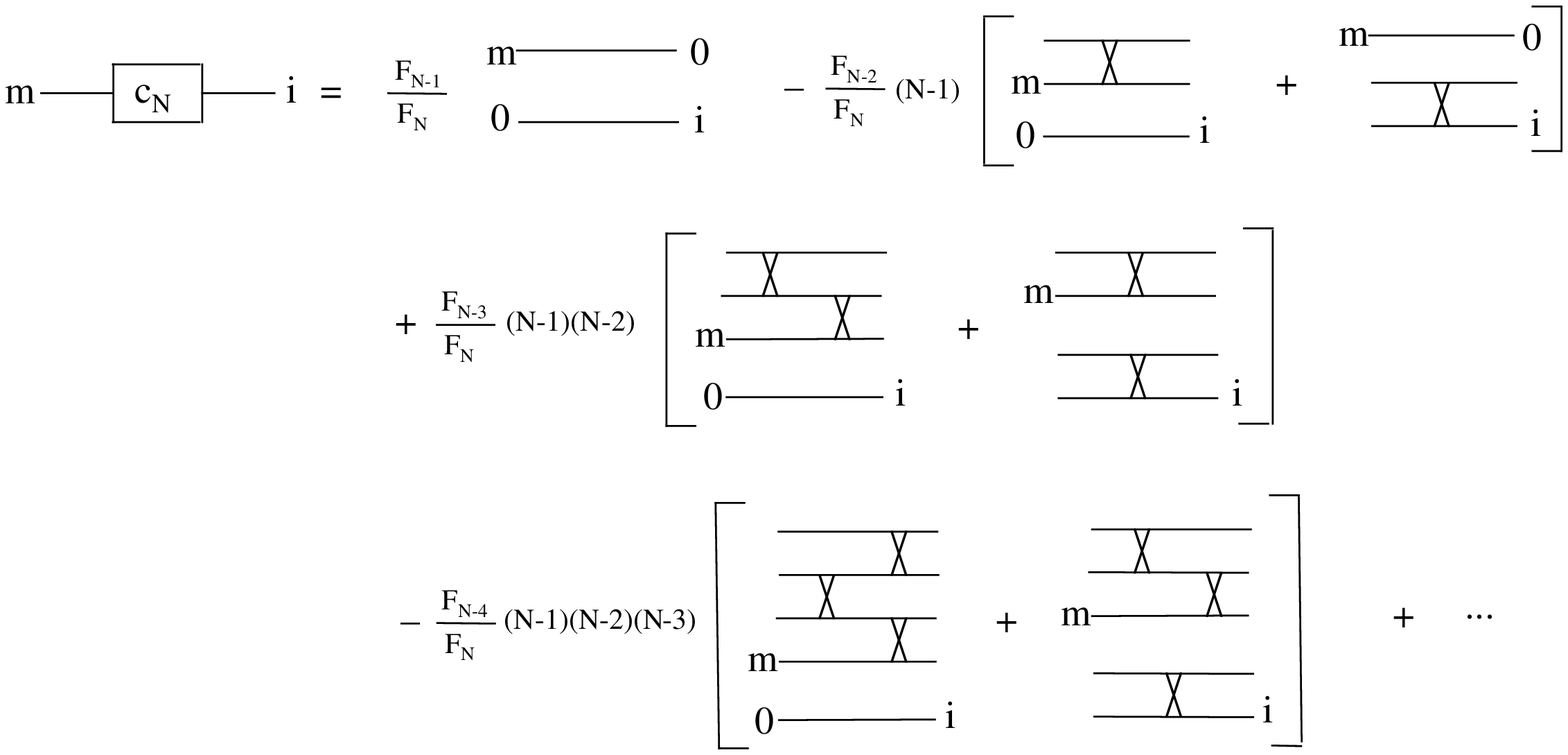}}}
\caption{\footnotesize{Pauli diagrams for $c_N(m,i)$ defined in eq.\ (49). This
$c_N(m,i)$ appears instead of $b_N(m,i)$ when we want to write
$A_N(m,i)$ in terms of $A_{N-2}(m,j)$ and not in terms of 
$A_{N-2}(n,i)$. As $b_N(m,i)$, $c_N(m,i)$ is made of disconnected
diagrams. At each order, it contains two terms, one with a
$\delta_{i0}$ factor, the other with a $\lambda_{000i}$ factor.}}
\end{figure}

\clearpage

\begin{figure}
\centerline{ \scalebox{0.7}{\includegraphics{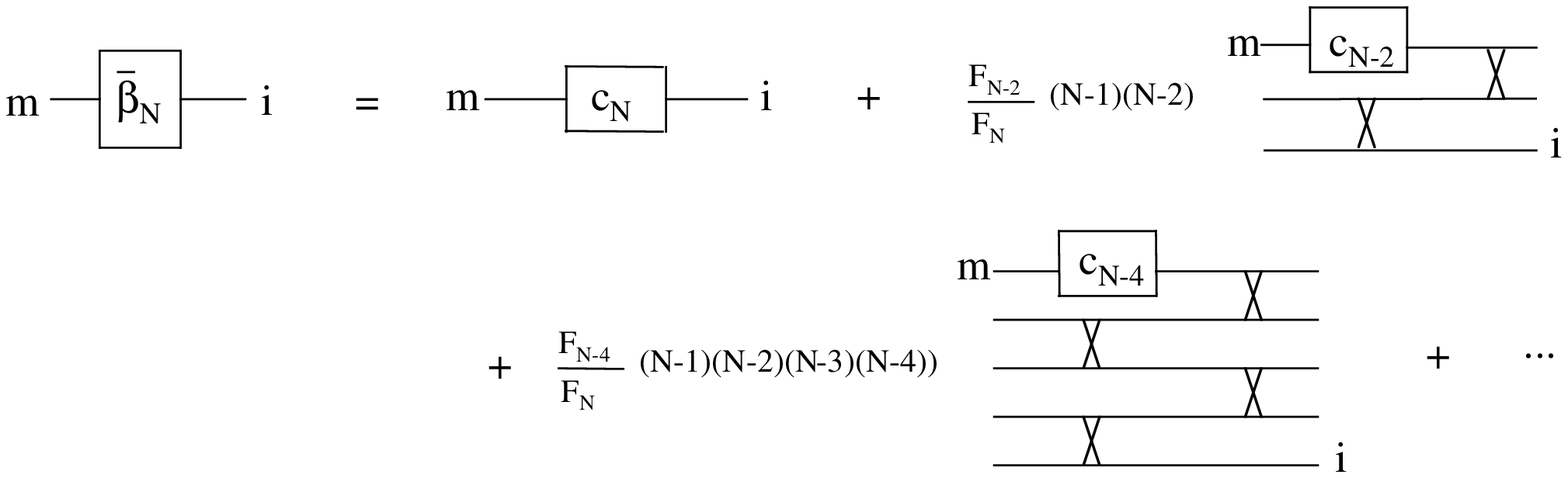}}}
\caption{\footnotesize{
Diagrammatic representation of the $\overline{\beta}_N(m,i)$
part of $A_N(m,i)$ which exists for $\v Q_m=\v Q_i=\v Q_0$ only,
as it appears when we use the recursion relation between
$A_N(m,i)$ and $A_{N-2}(m,j)$, instead of $A_{N-2}(n,i)$. This
diagrammatic representation corresponds to the iteration of eq.\
(52). By inserting the diagrams for $c_N(m,i)$, shown in fig.\
7, in this fig.\ 8, it is straightforward to see that the Pauli
diagrams for $\overline{\beta}_N(m,i)$ are identical to the ones
for $\beta_N(m,i)$ at any order in Pauli scatterings, so that 
$\overline{\beta}_N(m,i)=\beta_N(m,i)$.}}
\end{figure}

\clearpage

\begin{figure}
\centerline{ \scalebox{0.7}{\includegraphics{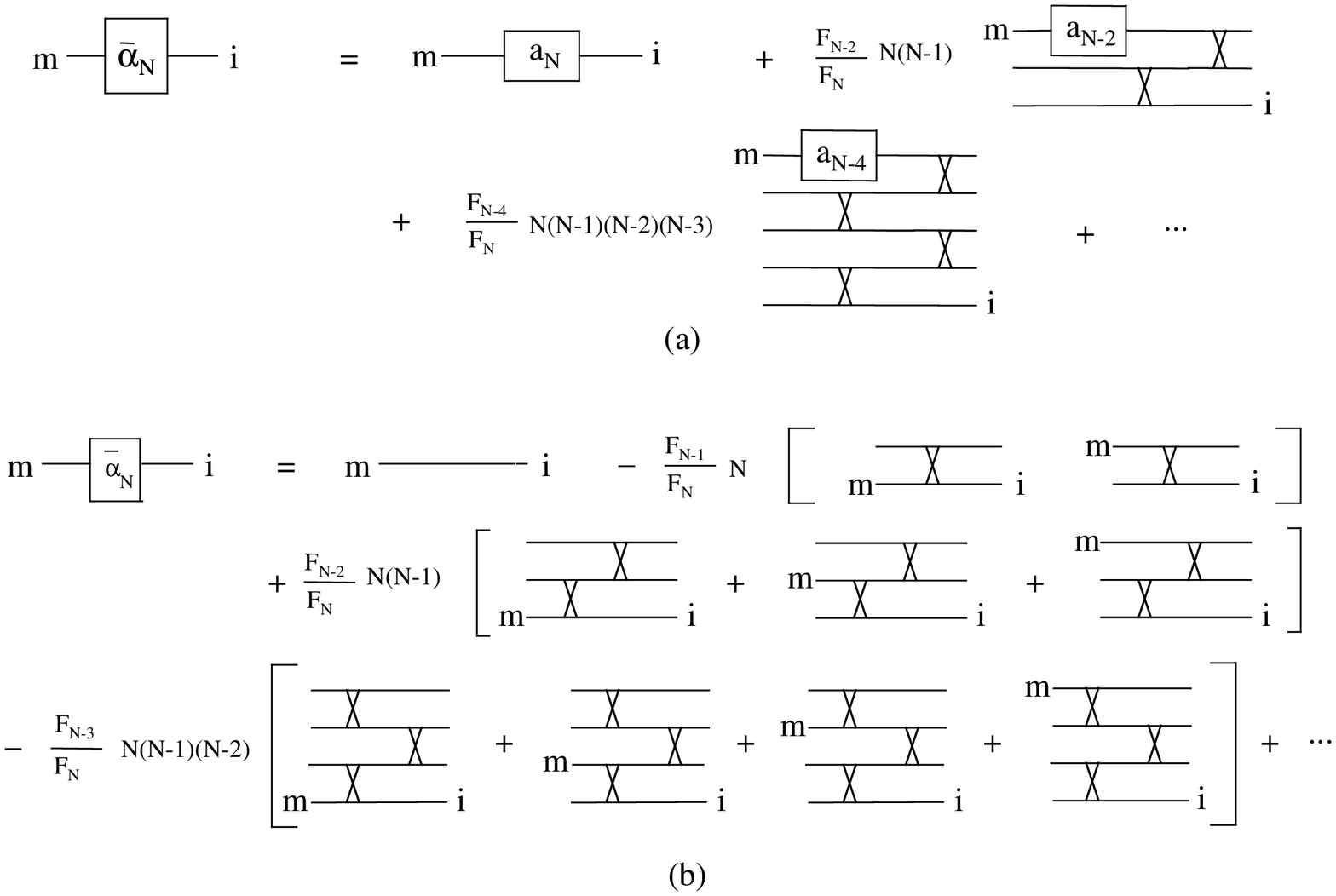}}}
\caption{\footnotesize{(a): Diagrammatic representation of the
$\overline{\alpha}_N(m,i)$ part of $A_N(m,i)$ which exists even
if $\v Q_m=\v Q_i\neq \v Q_0$, as it appears when we use the
recursion relation between $A_N(m,i)$ and $A_{N-2}(m,j)$. This
diagrammatic representation corresponds to the iteration of eq.\
(51). (b): Pauli diagrams for this $\overline{\alpha}_N(m,i)$ as
obtained by inserting the diagrams of fig.\ 3c for $a_N(m,i)$
into fig.\ 9a. These diagrams look like the ones for
$\alpha_N(m,i)$, shown in fig.\ 5c, except that the zigzags are
now left, right, left, \ldots, with $i$ staying at the right
bottom and $m$ moving at all possible positions on the left.
Since $\overline{\beta}_N(m,i)=\beta_N(m,i)$, we do have 
$\overline{\alpha}_N(m,i)=\alpha_N(m,i)$ for any $N$; so that
the two sets of Pauli diagrams for $\overline{\alpha}_N(m,i)$
and $\alpha_N(m,i)$ must actually correspond to the same
quantity, at any order in Pauli scatterings. This will be proved
later on.}}
\end{figure}

\clearpage

\begin{figure}
\centerline{ \scalebox{0.7}{\includegraphics{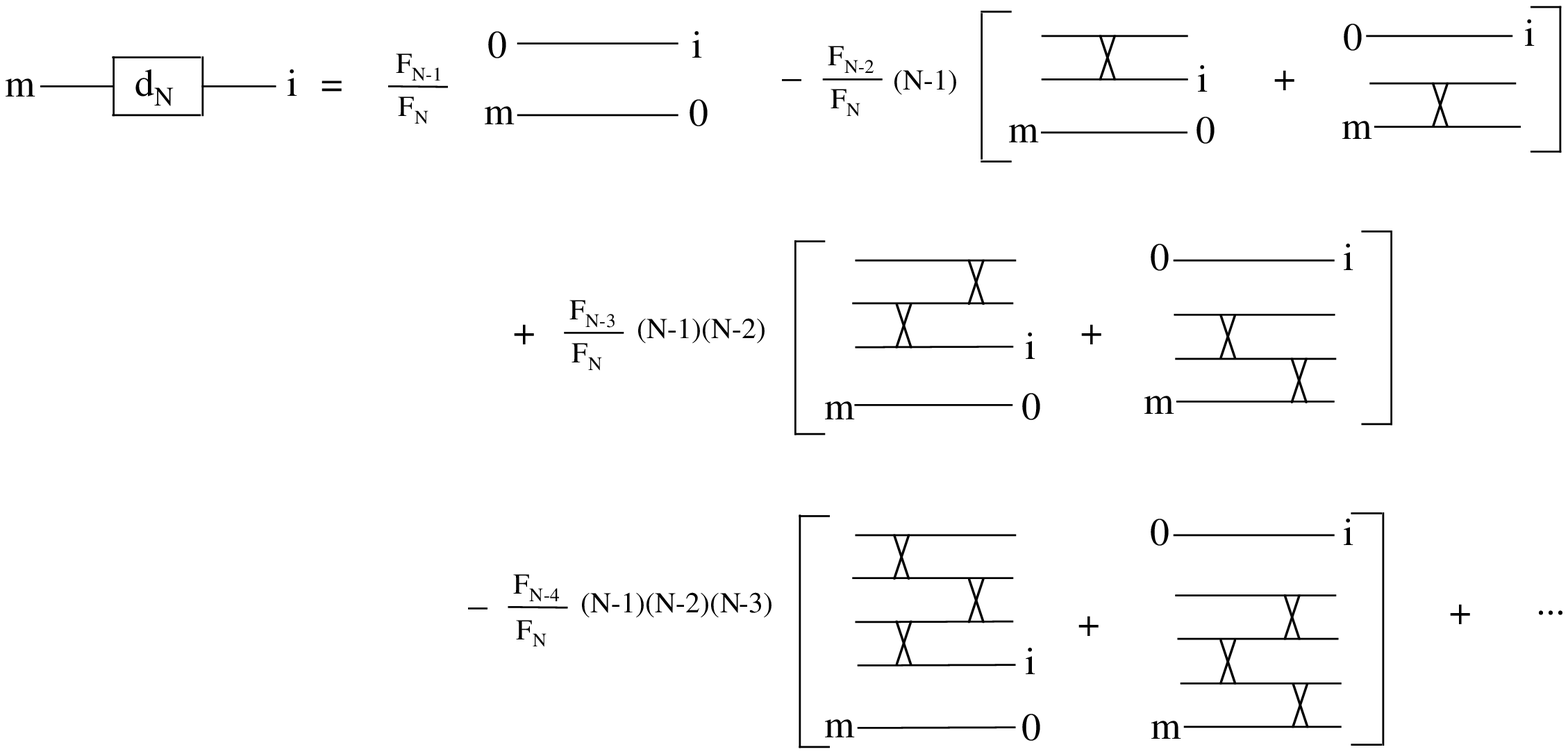}}}
\caption{\footnotesize{
Pauli diagrams for $d_N(m,i)$ defined in eq.\ (55). This 
$d_N(m,i)$ appears instead of $b_N(m,i)$ when we write
$A_N(m,i)$ in terms of $A_{N-2}(n,j)$ and not in terms of
$A_{N-2}(n,i)$. As $b_N(m,i)$, $d_N(m,i)$ is made of
disconnected diagrams, with two terms at each order, one with a
$\delta_{m0}$ factor, the other with a $\delta_{0i}$ factor.
}}
\end{figure}

\clearpage

\begin{figure}
\centerline{ \scalebox{0.7}{\includegraphics{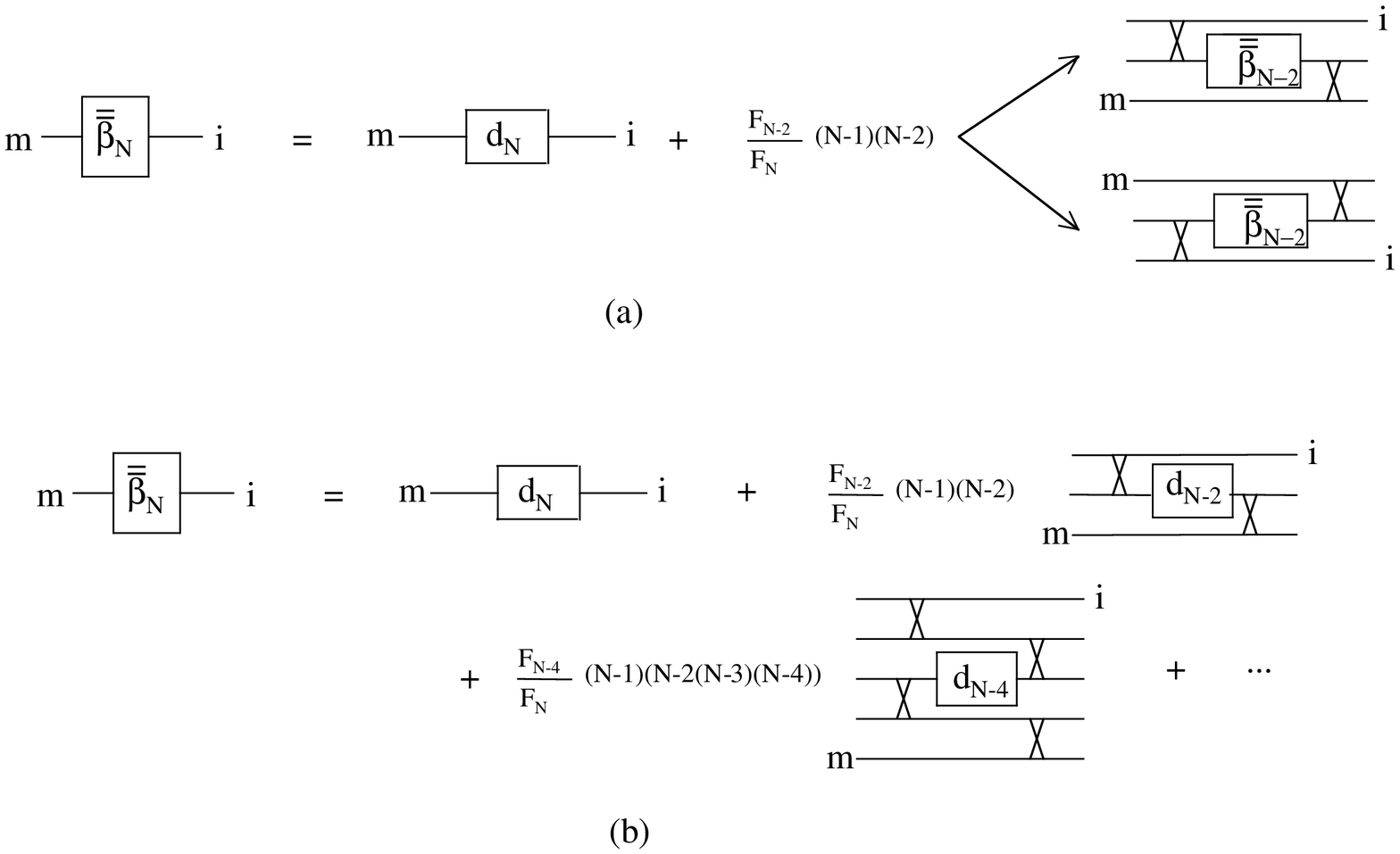}}}
\caption{\footnotesize{(a): The two possible representations of the same
recursion relation (58) for the $\overline{\overline{\beta}}_N(m,i)$ part
of $A_N(m,i)$ which exists for $\v Q_m=\v Q_i=\v Q_0$ only, as
it appears when we use the recursion relation between $A_N(m,i)$
and $A_{N-2}(n,j)$ instead of $A_{N-2}(n,i)$. These two
representations just result from a symmetry up-down which
follows from fig.\ 1c, since $\lambda_{mnij}=\lambda_{nmji}$.
(b): Iteration of this recursion relation. The two
representations of $\overline{\overline{\beta}}_N(m,i)$ shown in
fig.\ 11a have been alternatively used in order to avoid the
crossings of the exciton lines.
If, in this fig.\ 11b, we insert the diagrams of fig.\ 10 for
$d_N(m,i)$, we immediately find that the Pauli diagrams for
$\overline{\overline{\beta}}_N(m,i)$ are exactly those of
$\beta_N(m,i)$, so that $\overline{\overline{\beta}}_N(m,i)=
\beta_N(m,i)$.}}
\end{figure}

\clearpage

\begin{figure}
\centerline{ \scalebox{0.6}{\includegraphics{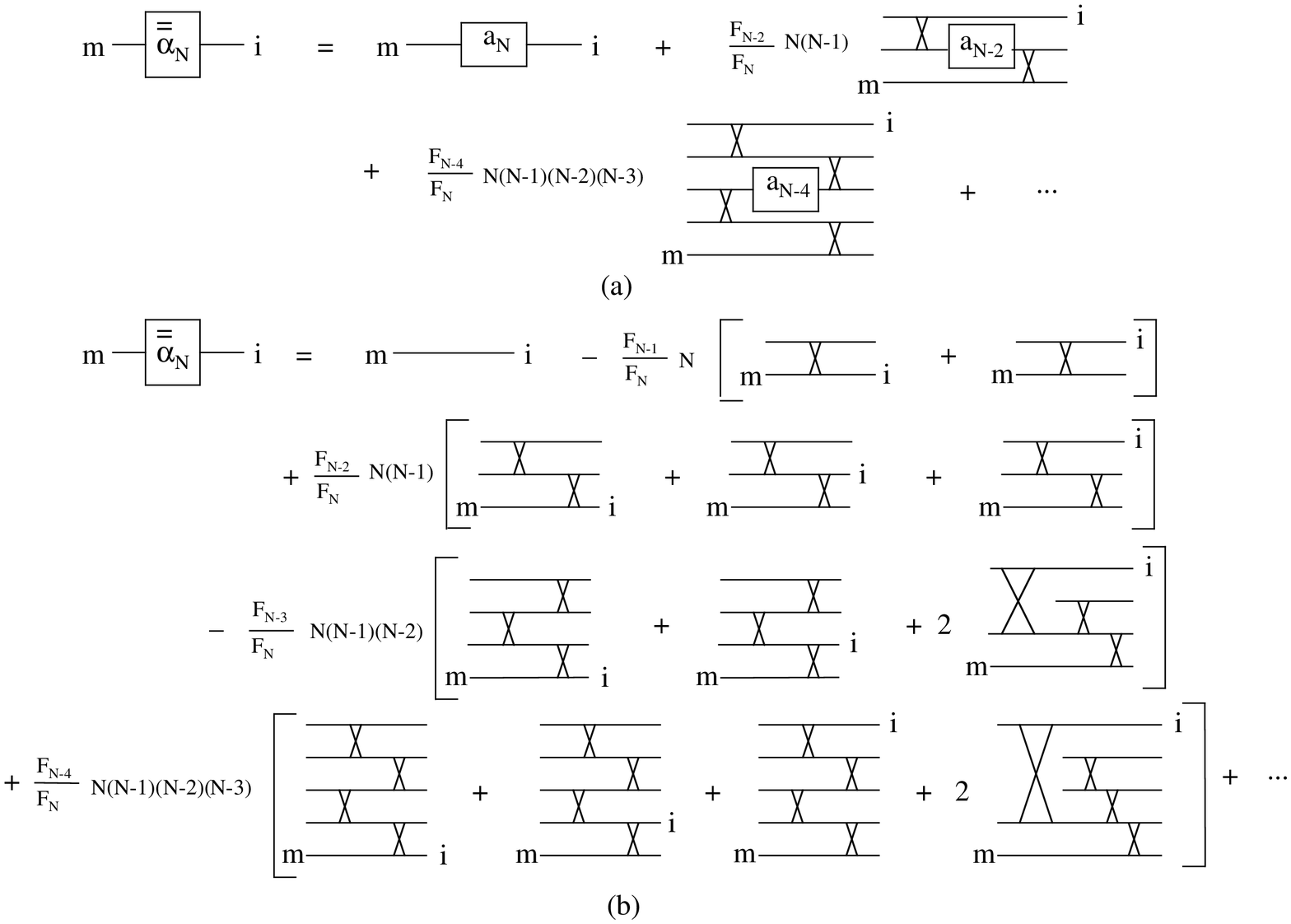}}}
\caption{\footnotesize{(a): Diagrammatic representation of the
$\overline{\overline {\alpha}}_N(m,i)$ part of $A_N(m,i)$ which exists even if
$\v Q_m=\v Q_i\neq\v Q_0$, as it appears when we use the recursion
relation between $A_N(m,i)$ and $A_{N-2}(n,j)$. This
diagrammatic representation corresponds to the iteration of eq.\
(57).
(b): Pauli diagrams for this $\overline{\overline
{\alpha}}_N(m,i)$ as obtained by inserting the diagrams of fig.\
3b for $a_N(m,i)$ into fig.\ 12a. The zeroth, first and second
order Pauli diagrams of $\overline{\overline
{\alpha}}_N(m,i)$ are identical to the ones of
$\alpha_N(m,i)$ shown in fig.\ 5c. On the opposite, at higher
orders, these Pauli diagrams become more and more different.
They however have to represent exactly the same quantity, since 
$\overline{\overline{\beta}}_N(m,i)=\beta_N(m,i)$, so that we do
have
$\overline{\overline{\alpha}}_N(m,i)=\alpha_N(m,i)$ for
\emph{any} $N$.}}
\end{figure}

\clearpage

\begin{figure}
\centerline{ \scalebox{0.8}{\includegraphics{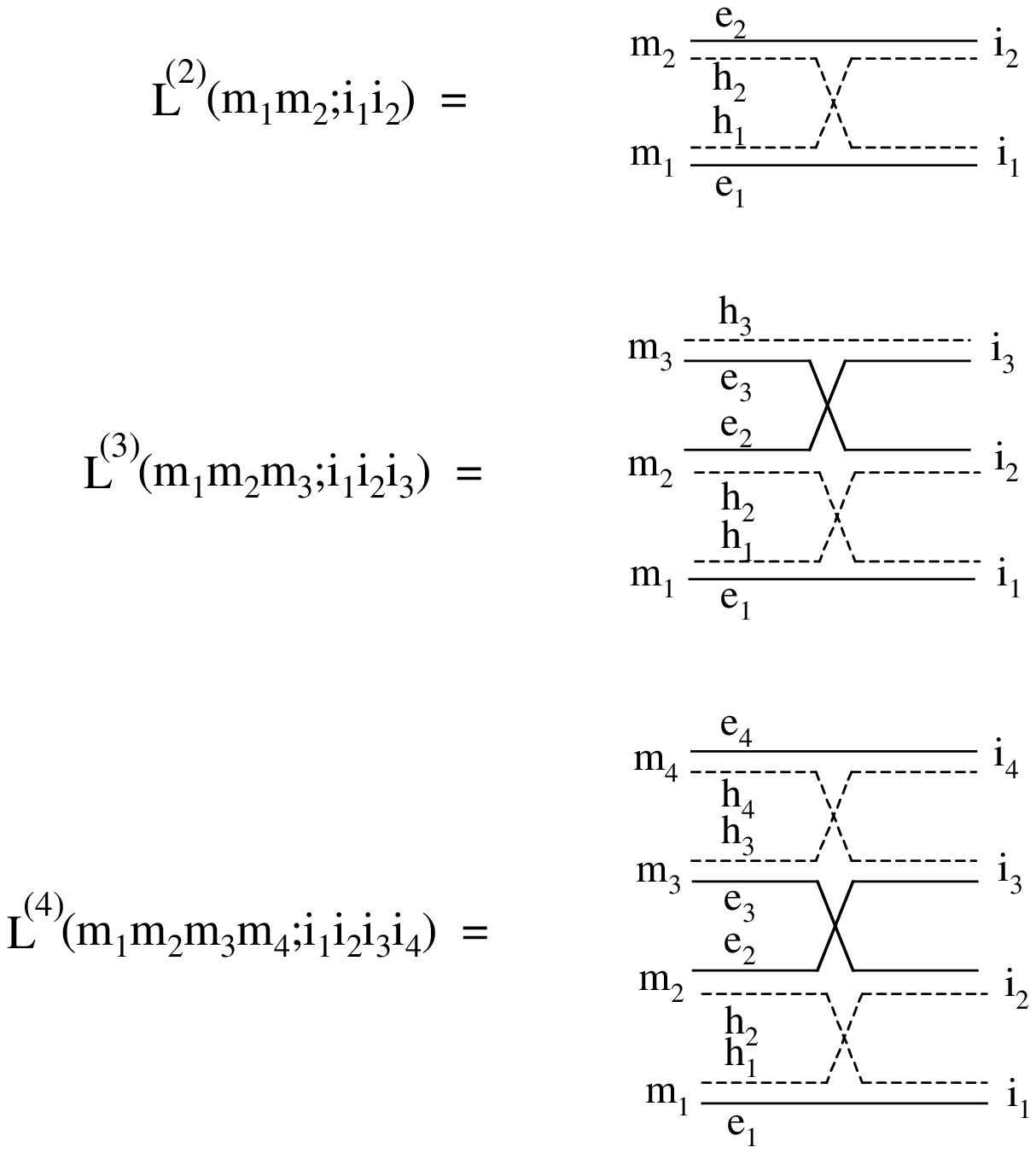}}}
\caption{\footnotesize{``Exchange skeletons'' between 2,3 and 4 excitons, as
defined in eqs.\ (59-61).}}
\end{figure}

\clearpage

\begin{figure}
\centerline{ \scalebox{0.8}{\includegraphics{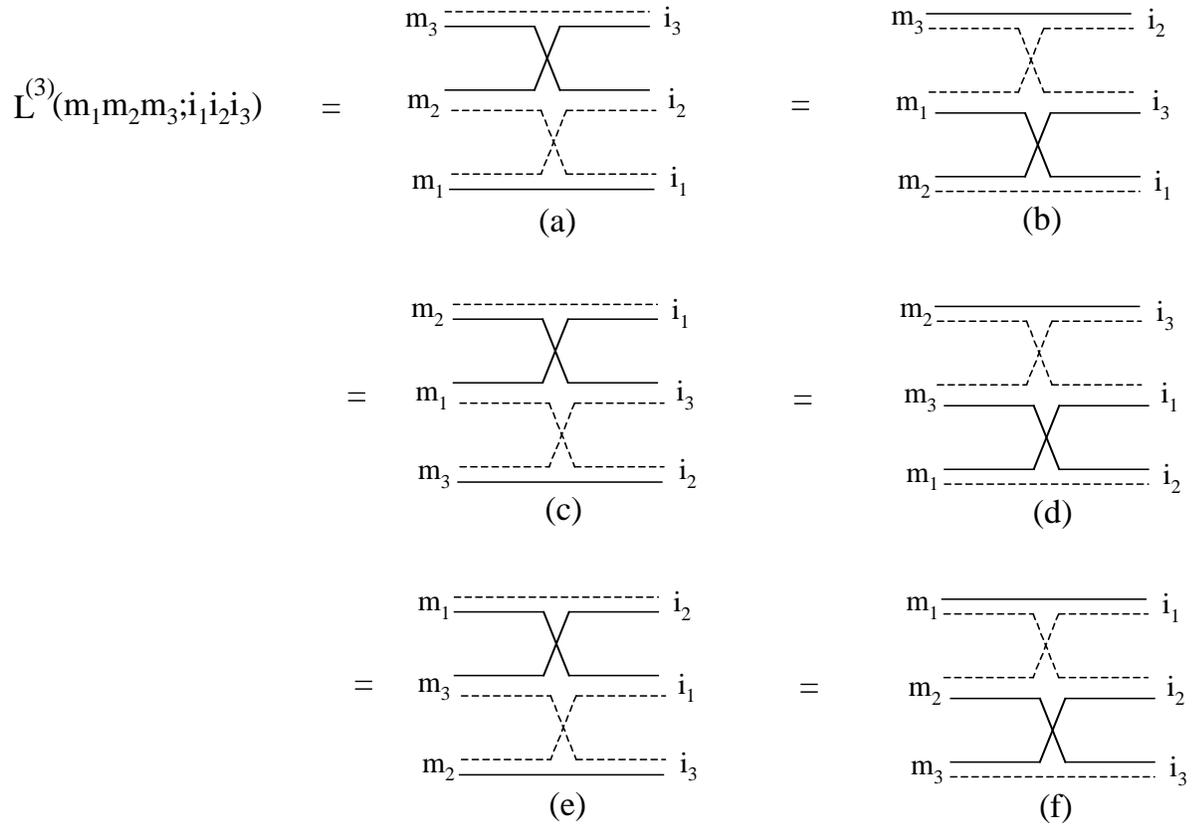}}}
\caption{\footnotesize{
Various possible representations of the ``exchange skeleton''
between three excitons. Note that, in all of them, the $m_1$ 
exciton is connected to the $i_1$ exciton by an electron line,
because these excitons have the same electron, while it is
connected to the $i_2$ exciton by a hole line, because they have
the same hole. And similarly for the other excitons.}}
\end{figure}

\clearpage

\begin{figure}
\centerline{ \scalebox{0.8}{\includegraphics{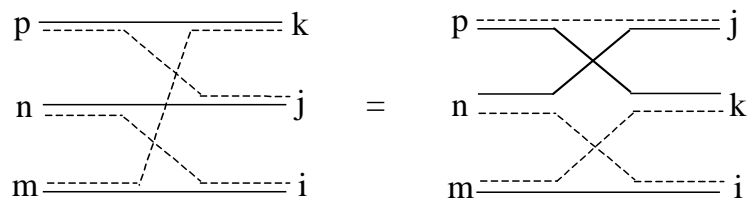}}}
\caption{\footnotesize{A possible carrier exchange between the $(m,n,p)$ and
$(i,j,k)$ excitons redrawn using an ``exchange skeleton'' between
three excitons.}}
\end{figure}

\clearpage

\begin{figure}
\centerline{ \scalebox{0.8}{\includegraphics{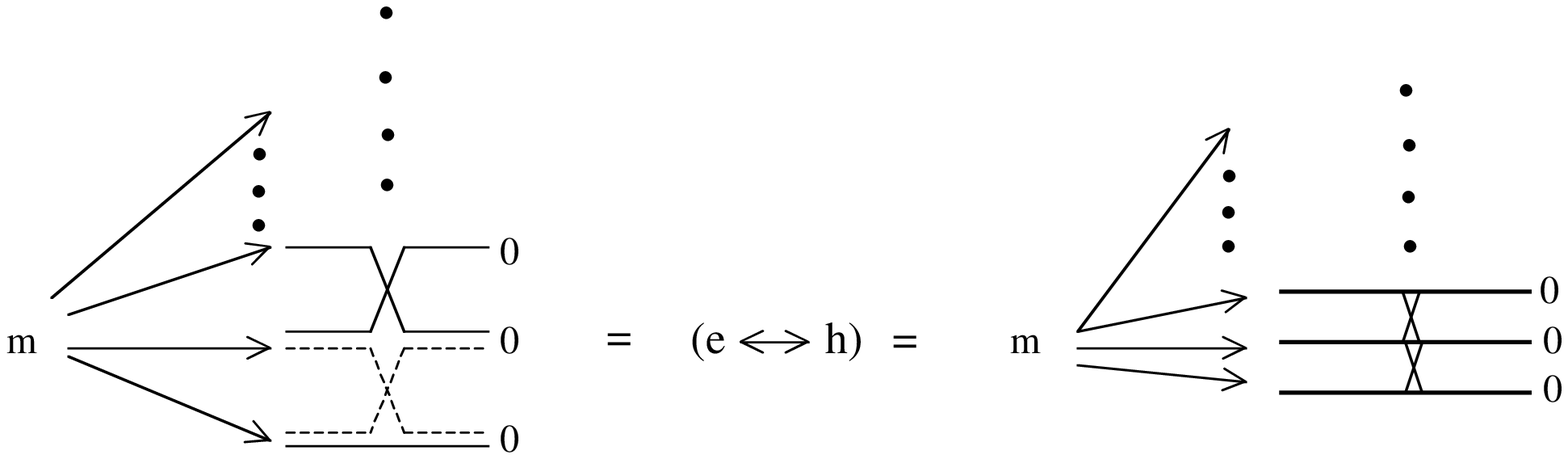}}}
\caption{\footnotesize{When all the indices of an exchange skeleton are equal,
except one, the position of this $m$ index can be in any
place. This exchange skeleton actually represents \emph{all}
Pauli diagrams with $m$ at any place on the left. The relative
positions of the crosses are also unimportant, due to possible
``slidings'' of the carrier exchanges, as shown in the next fig.\ 17
for the particular case of three excitons.}}
\end{figure}

\clearpage

\begin{figure}
\centerline{ \scalebox{0.75}{\includegraphics{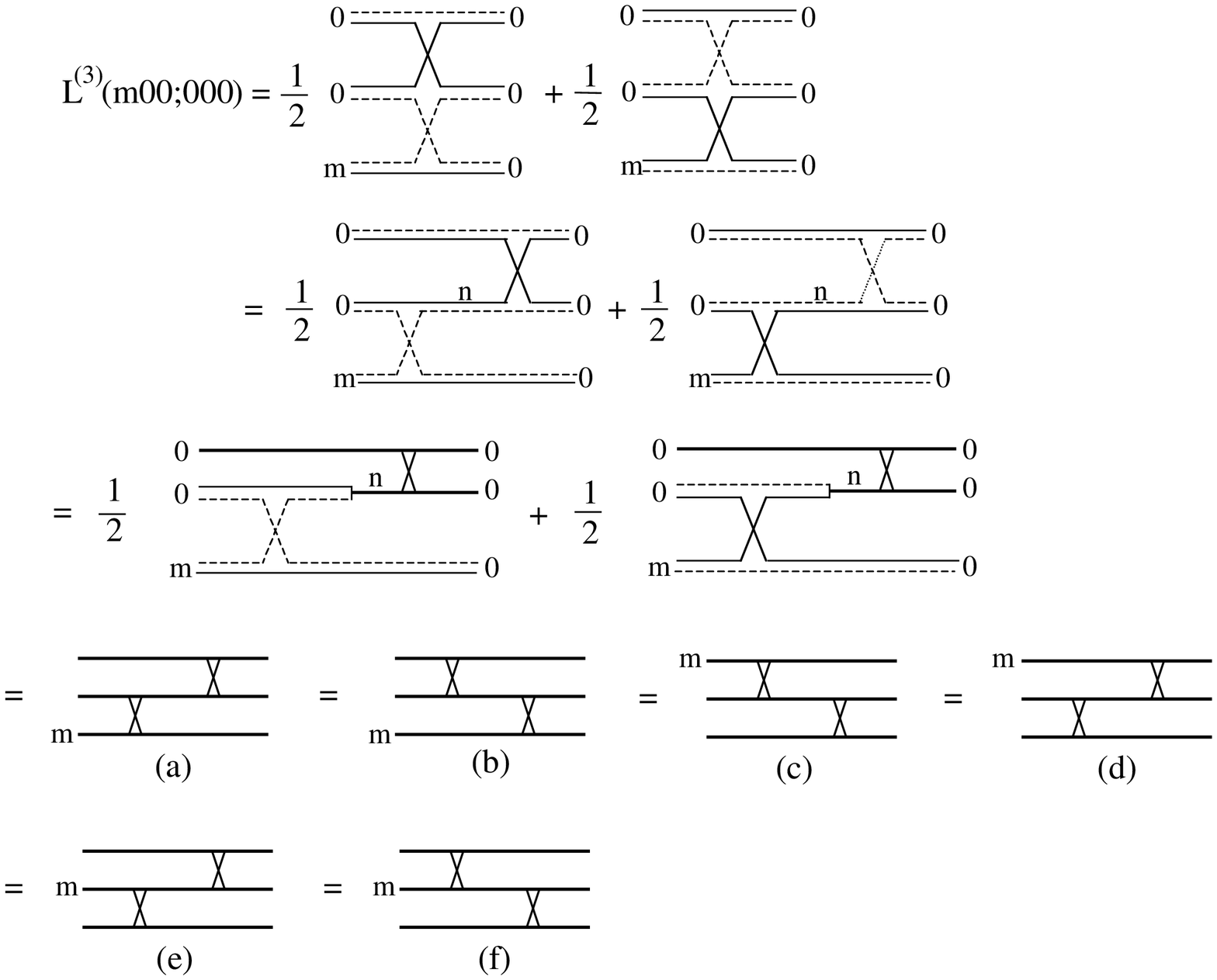}}}
\caption{\footnotesize{The exchange skeleton for three excitons
$L^{(3)}(m00;000)$, with all excitons except one in the same state, corresponds
to one of the two equivalent diagrams of the first line, so that it is also
half their sum. In the second line, we have just ``slided'' the carrier
exchanges. We then use the fact that, when the two indices on one side
of a Pauli scattering are equal, this Pauli scattering, here
$\lambda_{n000}$, corresponds either to an electron exchange or to a
hole exchange. Finally, we use the representation of fig.\ 1a for
$\lambda_{m00n}$ to get the Pauli diagram  shown in (a).
If we slide the carrier exchanges the other way, we get the
Pauli diagram (b).
The diagram (c) is obtained from (a)
by a symmetry up-down which follows from
$\lambda_{mnij}=\lambda_{nmji}$; and similarly for diagram 
(d) starting from (b).
In the same way, the diagram (e) follows from 
(a) due to $\lambda_{m00n}=\lambda_{0m0n}$, while the
diagram (f) follows from (c) for the same reason.
All this shows that the Pauli diagrams with only one index
different from $0$, correspond to the same exchange skeleton,
whatever the positions of $m$ and the Pauli scatterings are.}}
\end{figure}

\clearpage

\begin{figure}
\centerline{ \scalebox{0.5}{\includegraphics{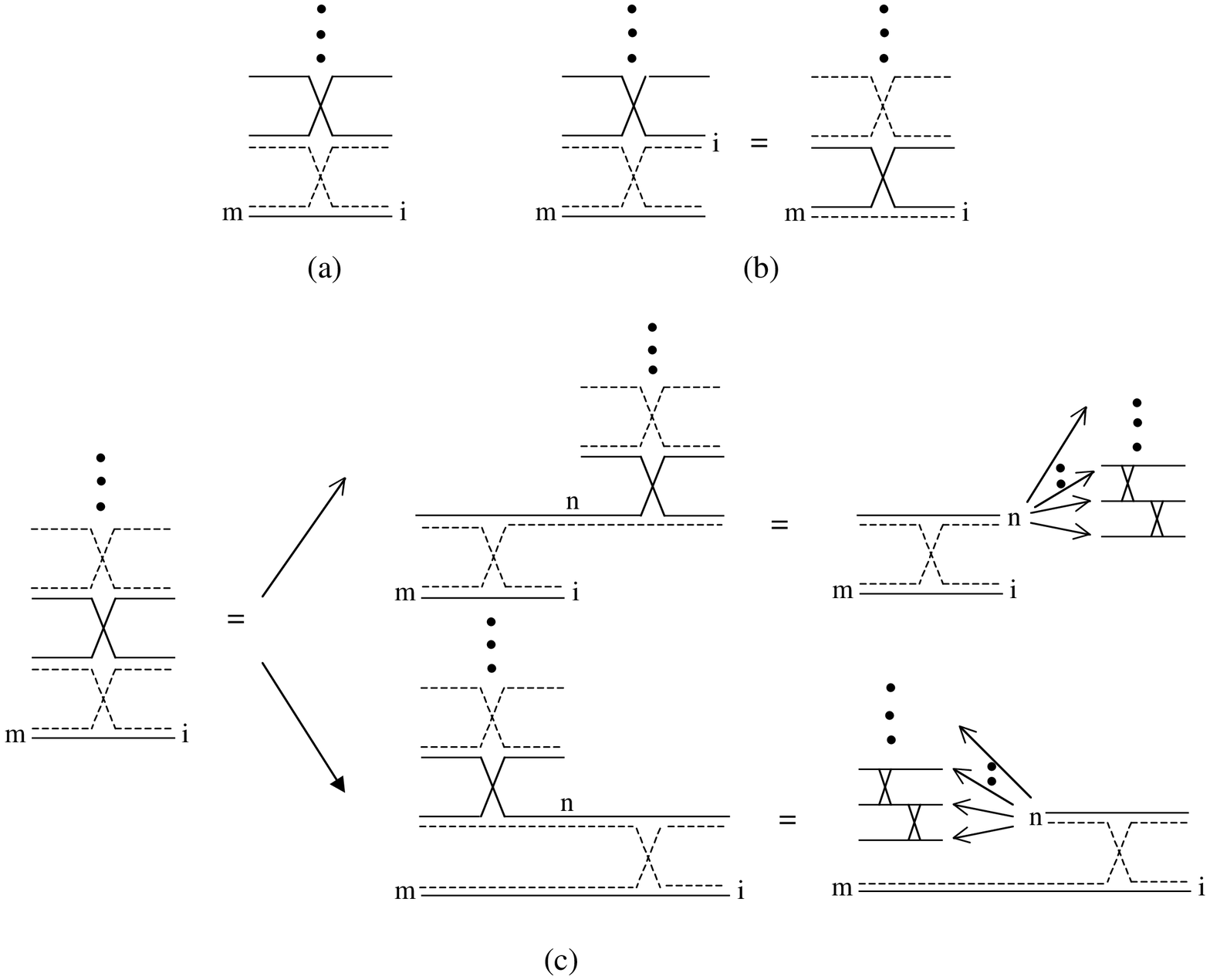}}}
\centerline{ \scalebox{0.5}{\includegraphics{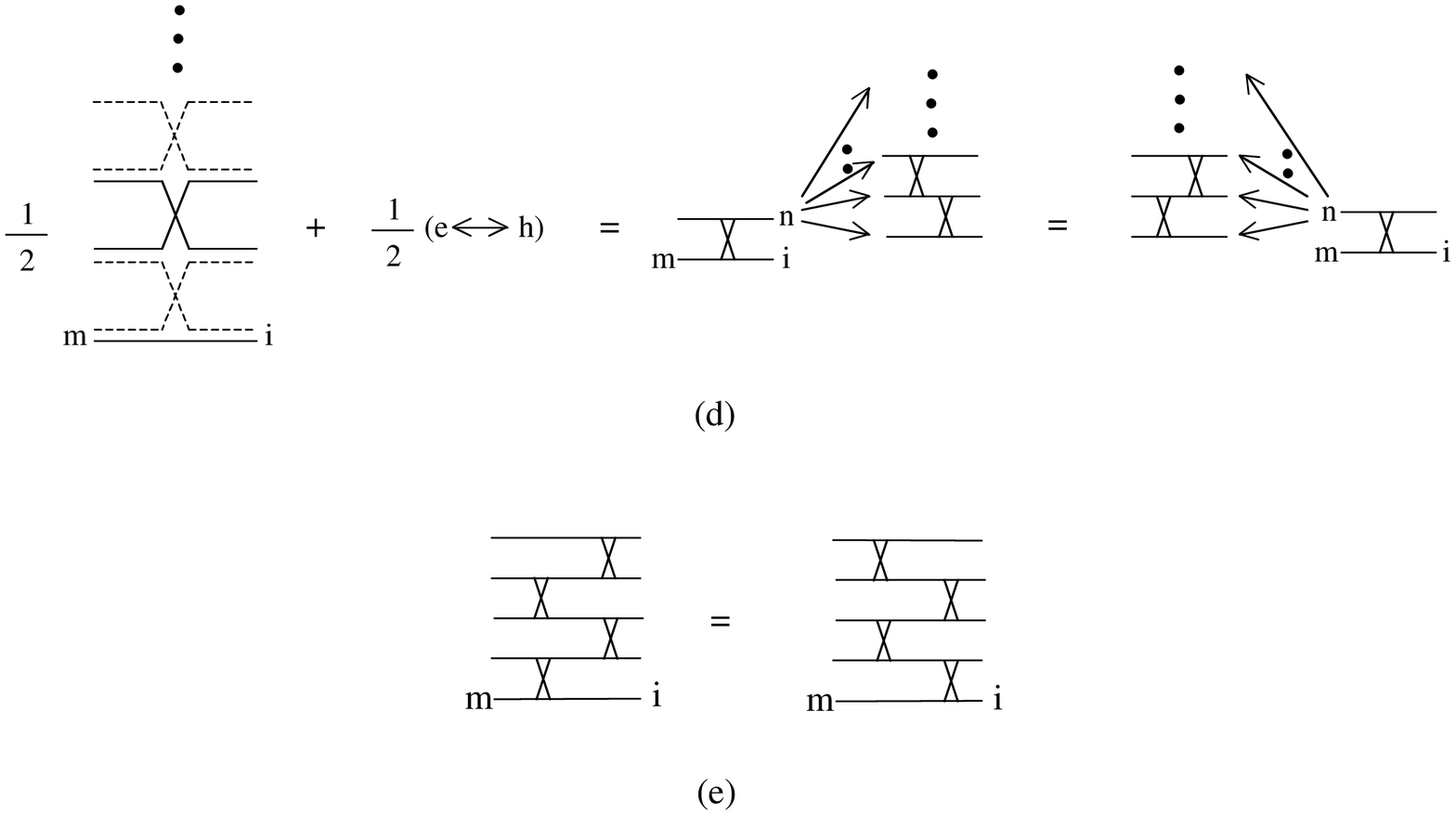}}}
\caption{\footnotesize{(a,b): Exchange skeletons for which the $m$ and $i$
excitons have the same electron (a) and the same hole (b).
(c): When all the other indices are $0$, we can slide the carrier
exchange one way or the other to identify an exchange skeleton
for which all the indices are $0$ except one; so that it
represents Pauli diagrams with the $n$ index and the crosses for Pauli
scatterings at any place, as shown in fig.\ 16.
(d,e): If we now add the electron exchange and the hole exchange
to restore a full $\lambda_{m0in}$ or $\lambda_{mni0}$ Pauli
scattering, we easily identify the set of Pauli diagrams which
represent the same quantity. This in particular shows that
the two Pauli diagrams shown in (e), which appear in the
two possible ways to calculate $a_N(m,i)$, namely the diagrams
of figs.\ 3b and 3c, are indeed equal.}}
\end{figure}

\clearpage

\begin{figure}
\centerline{ \scalebox{0.7}{\includegraphics{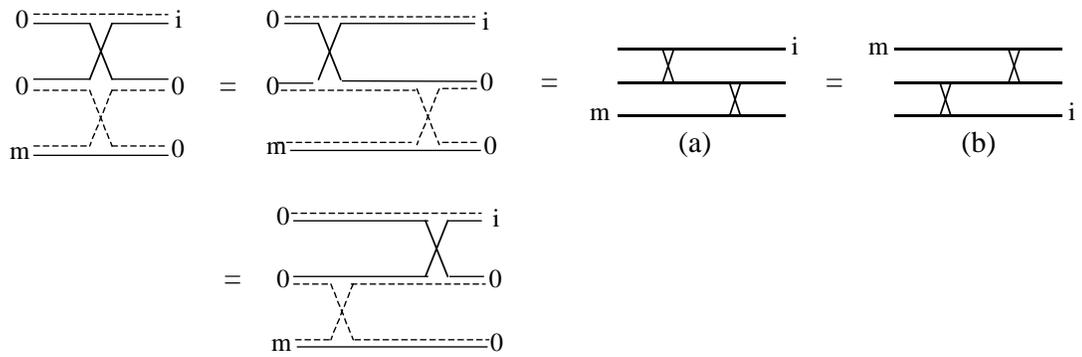}}}
\caption{\footnotesize{In the case of three excitons, there is only one
exchange
skeleton in which $m$ and $i$ have no common carrier. If the
other excitons are $0$ excitons, we can slide the electron
exchange to the left to make appearing two Pauli scatterings
with two equal indices on one side. Using figs.\ 1d and 1e, we
then find that this exchange skeleton corresponds to the Pauli
diagram (a). A symmetry up-down leads to the Pauli diagram
(b). Note that sliding the electron exchange to the right
would be of no help.}}
\end{figure}

\clearpage

\begin{figure}
\centerline{ \scalebox{0.5}{\includegraphics{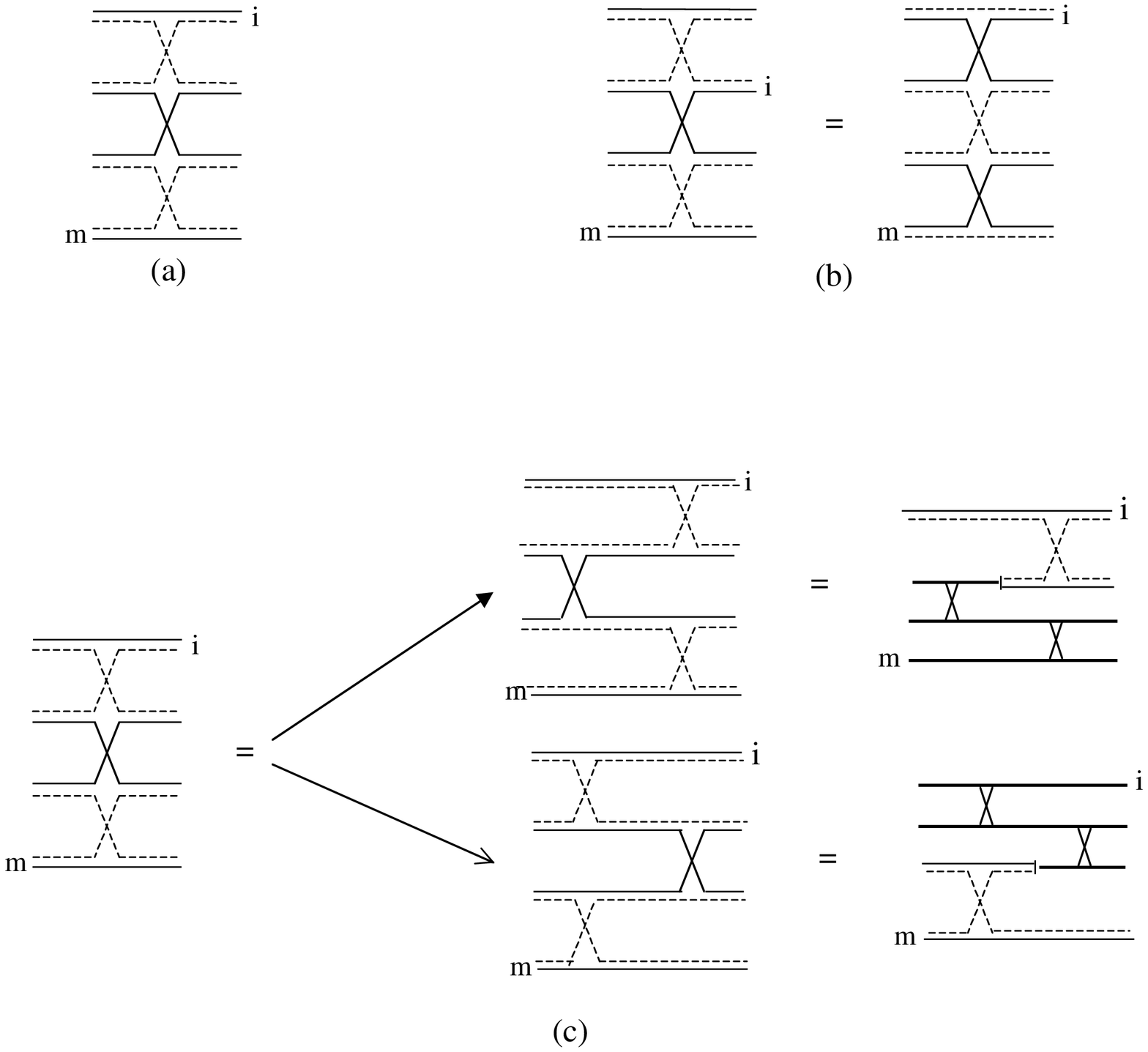}}}
\centerline{ \scalebox{0.5}{\includegraphics{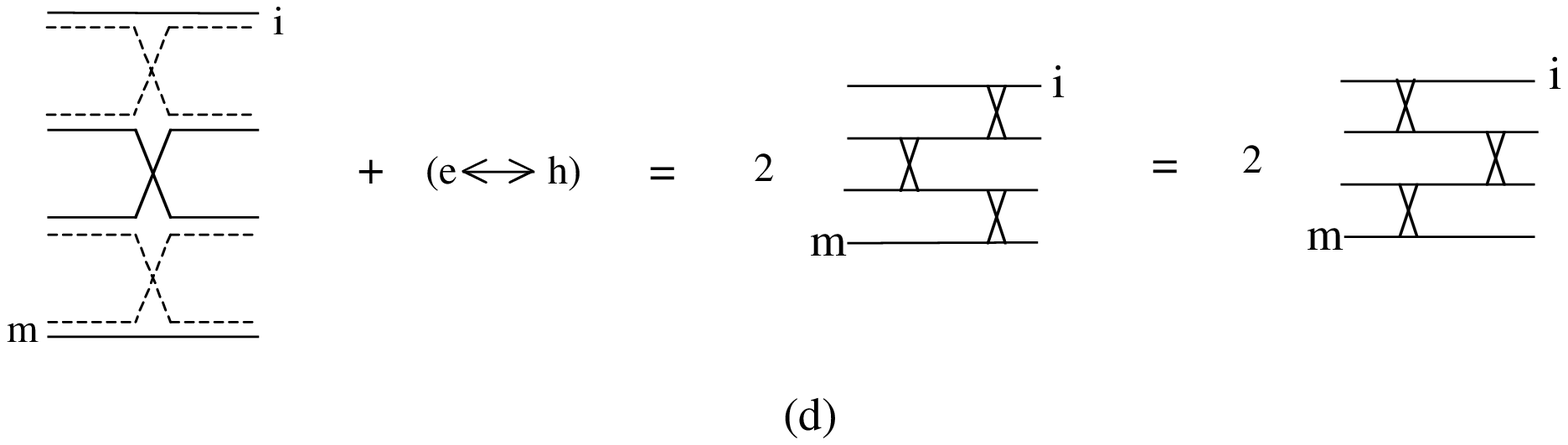}}}
\caption{\footnotesize{
(a,b): In the case of four excitons, there are two different
exchange skeletons in which the $m$ and $i$ excitons have no
common carrier, as shown in (a) and (b). They
correspond to exchange the role played by the electrons
and holes.
(c): Starting from the exchange skeleton (a), we can slide the
carrier exchanges one way or the other, to make appearing Pauli
scatterings with two identical excitons $0$ on one side.
(d): By combining the two exchange skeletons (a) and (b), we
can restore a full $\lambda_{n00i}$ or $\lambda_{m00n}$ to get
the two Pauli diagrams (d). They just differ by the
order of the zigzags, right, left, right or left, right, left.}}
\end{figure}

\clearpage

\begin{figure}
\centerline{ \scalebox{0.5}{\includegraphics{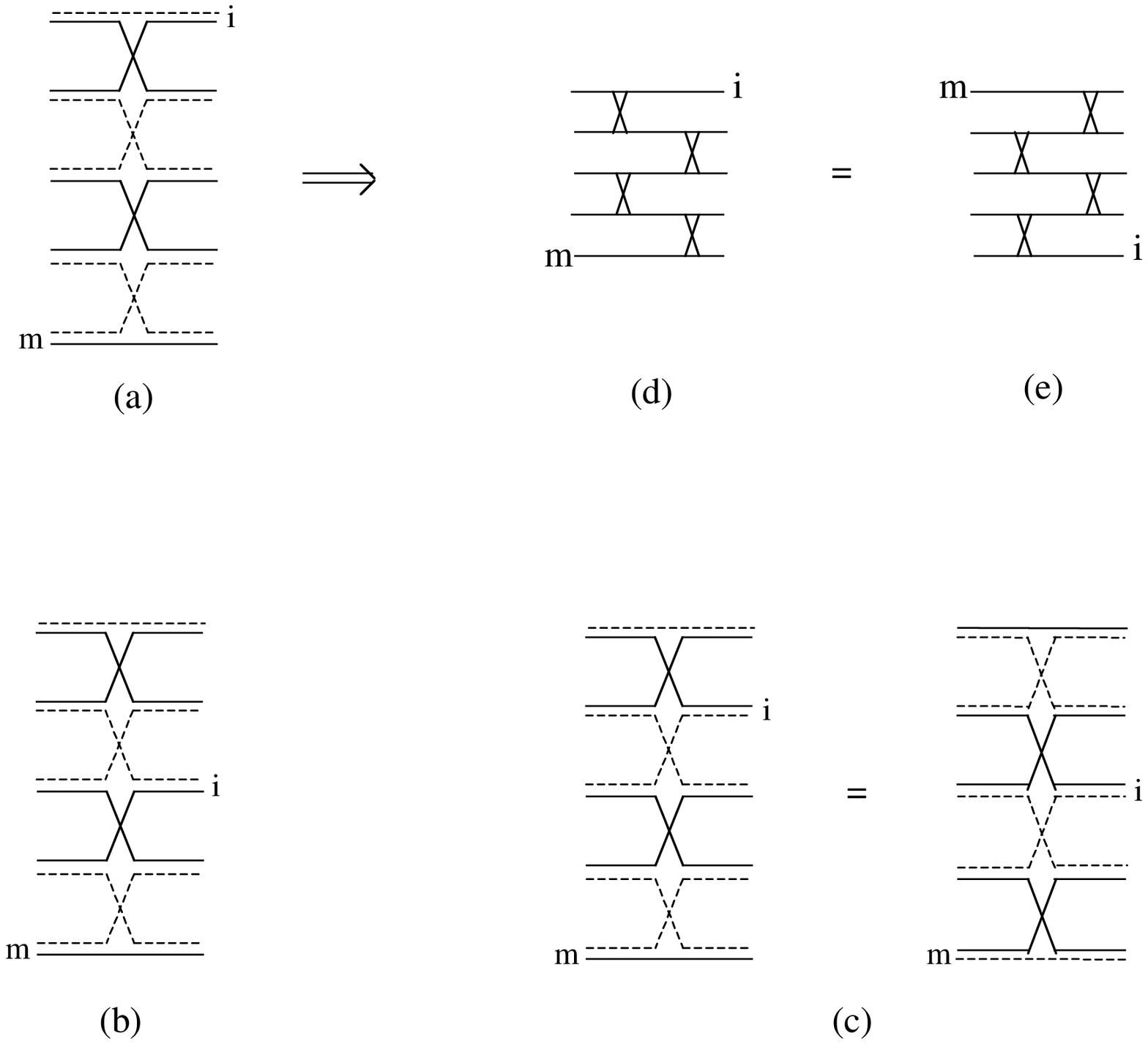}}}
\centerline{ \scalebox{0.5}{\includegraphics{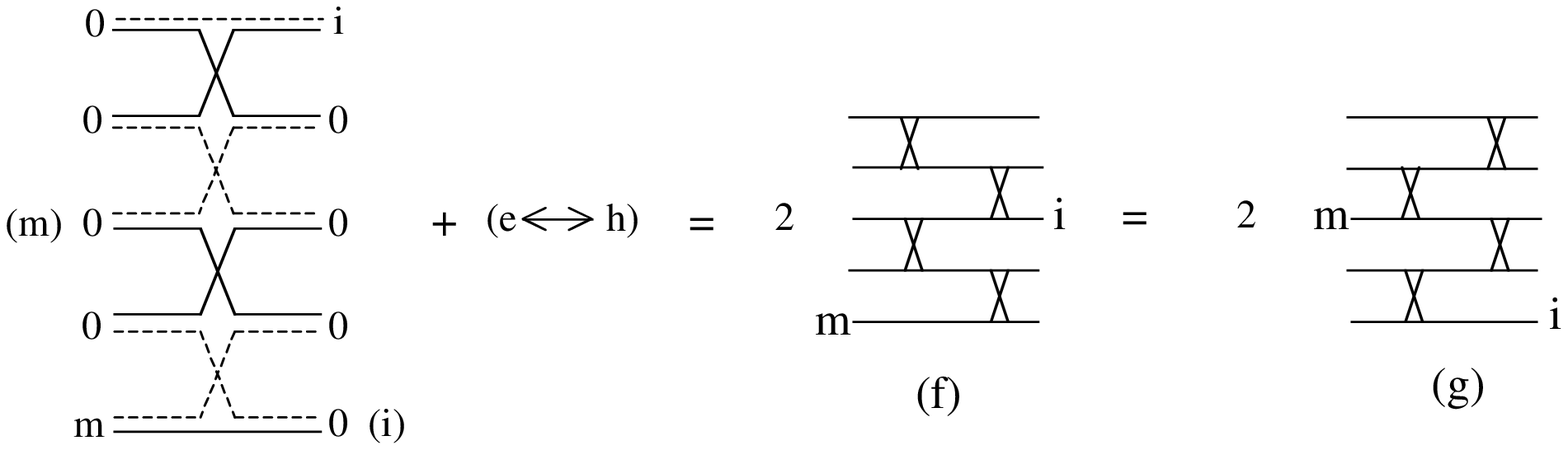}}}
\caption{\footnotesize{
In the case of five excitons, there are three exchange skeletons
in which $m$ and $"$ have no common carrier. They are shown in figures (a,b,c).
By sliding the carrier exchanges of the exchange skeleton (a)
with all the other indices equal to $0$, we immediately get the Pauli
diagram (d). The diagram (e) follows from an up-down
symmetry.
If we now add the exchange skeletons (b) and (c) --- which
just correspond to exchange the roles played by the electrons and
holes --- and we again slide the carrier exchanges to make
appearing Pauli scatterings with two indices on one side equal to
$0$, we get the Pauli diagram (f).
In the same way, it is possible to obtain the diagram (g) from
the same exchange skeletons, by noting that, when all the
other indices are equal, it is possible to exchange the
positions of the $m$ and $i$ indices in these skeletons.}}
\end{figure}

\clearpage

\begin{figure}
\centerline{ \scalebox{0.7}{\includegraphics{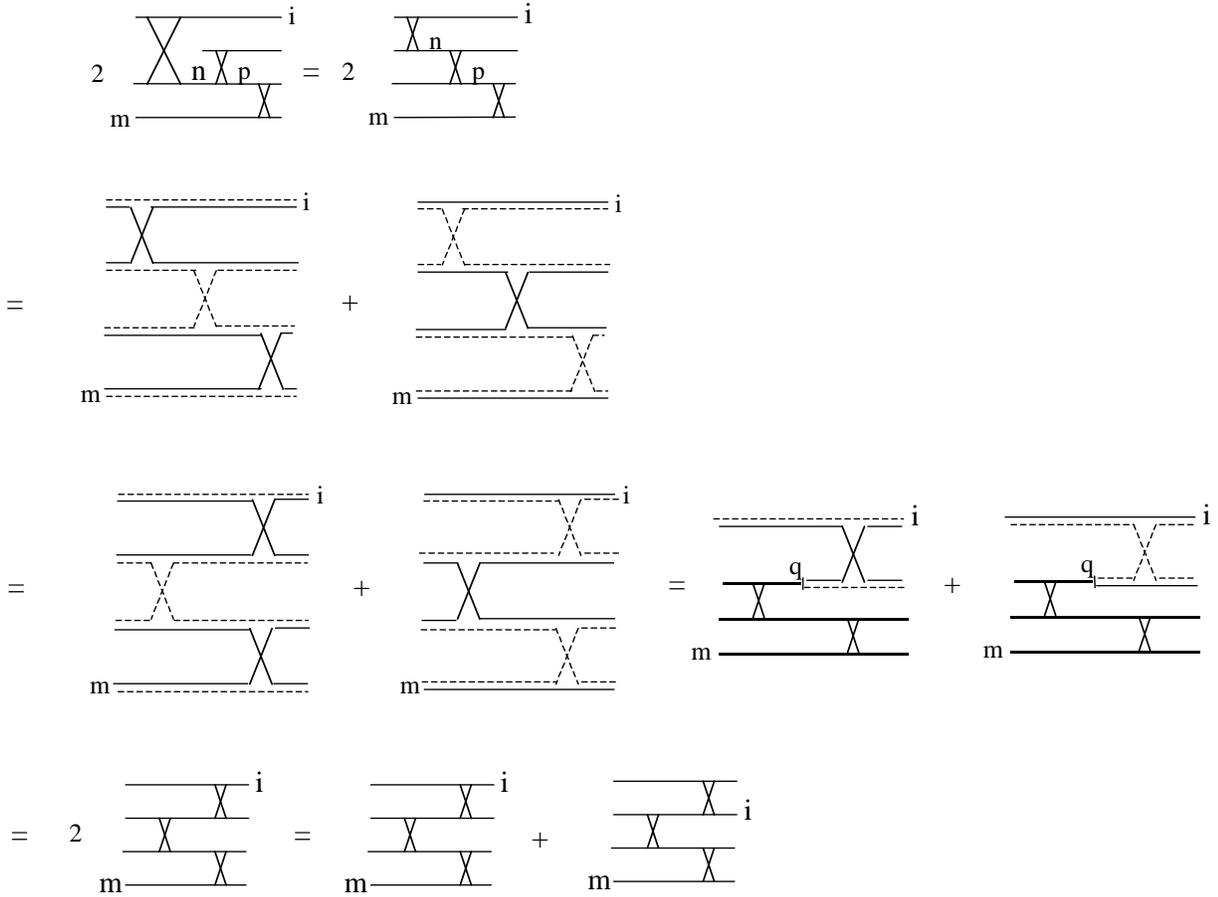}}}
\caption{\footnotesize{In this figure, we show how to transform the last third
order
Pauli diagram of $\overline{\overline{\alpha}}_N(m,i)$ appearing
in fig.\ 12b, into the two missing zigzag diagrams of fig.\ 5c
which enter $\alpha_N(m,i)$, as reproduced at the bottom of this
figure.
We first use the fact that $\lambda_{n0p0}=\lambda_{0np0}$ to
get an equivalent representation of this diagram.
We then note that the upper and lower crosses have two identical
excitons on one side. By using figs.\ 1d and 1e for the two
upper crosses, and fig.\ 1a and 1b for the middle cross, we get
the two diagrams of the second line of this fig.\ 22. The
diagrams of the next line just follow by sliding the carrier
exchanges. In them, we now identify other Pauli scatterings with
two identical excitons on one side. These two diagrams can be
used to restore the full Pauli scattering $\lambda_{q00i}$. The
sum of the last two zigzag diagrams simply results from
$\lambda_{q00i}=\lambda_{q0i0}$.}}
\end{figure}

\clearpage

\begin{figure}
\centerline{ \scalebox{0.7}{\includegraphics{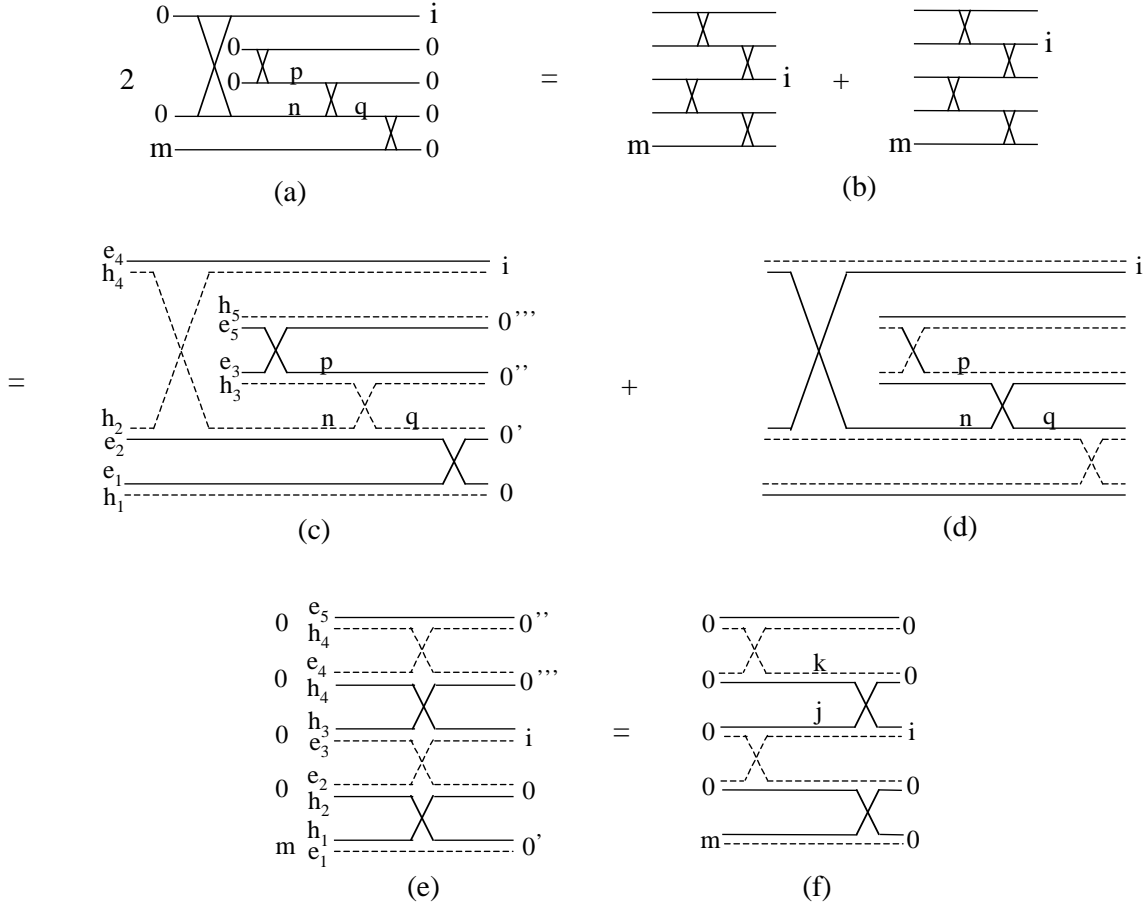}}}
\caption{\footnotesize{The transformation of the ugly fourth order Pauli
diagrams
appearing in $\overline{\overline{\alpha}}_N(m,i)$ and
reproduced in (a), into the two missing zigzag diagrams of
$\alpha_N(m,i)$ reproduced in (b), is actually quite subtle.
In diagram (a), all the Pauli scatterings except
$\lambda_{npq0}$ have two $0$ indices on one side. This allows
to transform twice the diagram (a) into the sum of diagrams (c)
and (d), using figs.\ 1a,1b and figs.\ 1d,1e.
Diagram (c) is nothing but diagram (e) as easy to see by
following the electron and hole lines, in order to check that the
excitons with identical electrons or holes are indeed connected.
We then slide the carrier exchange to make appearing a Pauli
scattering with two excitons $0$ on one side. They are all of
this type except the one between $(j,k)$ and $(i,0)$. Using
diagram (d), we can restore the full Pauli scattering
$\lambda_{jk0i}$. The two diagrams shown in (b) simply result
from $\lambda_{jk0i}=\lambda_{jki0}$.}}
\end{figure}

\end{document}